\documentclass[twocolumn,trackchanges]{63}
\usepackage{silence} 
\usepackage{natbib}
\usepackage{amsmath}
\usepackage{float}
\usepackage{times}
\usepackage{enumitem}
\usepackage{etoolbox}
\usepackage{graphicx}
\usepackage{amsmath}
\usepackage{bm}
\usepackage{graphics}
\usepackage{url}
\usepackage{epsfig}
\usepackage{wrapfig}
\usepackage{ulem}
\usepackage{verbatim}
\usepackage{float}
\usepackage{lastpage}

\def\lsim{\hbox{\rlap{\raise 0.425ex\hbox{$<$}}\lower 0.65ex\hbox{$\sim$}}}
\def\gsim{\hbox{\rlap{\raise 0.425ex\hbox{$>$}}\lower 0.65ex\hbox{$\sim$}}}

\def\arcdeg{\mbox{$^\circ$}}

\newcommand\N[1]{{\color{blue} \bf #1}} %N8

\shorttitle{Ultraviolet Spectroscopy of SN~2021yja}
\shortauthors{Vasylyev et al.}

\begin{document}

%***\alex\Are we still updating the light curve? How many months -- about 3? So, we should be able to tell whether it's Type II-P.
\title{Early-Time Ultraviolet Spectroscopy and Optical Follow-up Observations
of the Type IIP Supernova 2021yja}
\author[0000-0002-4951-8762]{Sergiy~S.Vasylyev}
\affiliation{Department of Astronomy, University of California, Berkeley, CA 94720-3411, USA}
\affiliation{Steven Nelson Graduate Fellow}
\author[0000-0003-3460-0103]{Alexei~V.~Filippenko}
%\author{Sergiy~S.~Vasylyev, Alexei~V.~Filippenko}
%\affiliation{Department of Astronomy, University of California, Berkeley, CA 94720-3411, USA}
\affiliation{Department of Astronomy, University of California, Berkeley, CA 94720-3411, USA}
%\affiliation{Miller Senior Fellow, Miller Institute for Basic Research in Science, University of California, Berkeley, CA 94720, USA}
\author[0000-0002-7941-5692]{Christian Vogl}
\affiliation{Max-Planck-Institut f\"ur Astrophysik, Karl-Schwarzschild-Str. 1, 85748 Garching, Germany}
\author{Thomas~G.~Brink}
\affiliation{Department of Astronomy, University of California, Berkeley, CA 94720-3411, USA}
\author[0000-0001-6272-5507]{Peter J. Brown}
\affiliation{Mitchell Institute for Fundamental Physics and Astronomy,\\ Department of Physics and Astronomy, Texas A\&M University \\
4242 TAMU, College Station, TX, USA}
\author{Thomas~de~Jaeger}
\affiliation{Institute for Astronomy, University of Hawaii, 2680 Woodlawn Drive, Honolulu, HI 96822, USA.}
\author[0000-0001-6685-0479]{Thomas Matheson}
\affiliation{NSF's National Optical-Infrared Astronomy Research Laboratory, 950 North Cherry Avenue, Tucson, AZ 85719, USA}
\author{Avishay Gal-Yam}
\affiliation{Department of Particle Physics and Astrophysics, Weizmann Institute of Science, 76100 Rehovot,
Israel}
\author{Paolo A. Mazzali}
\affiliation{Max-Planck-Institut f\"{u}r Astrophysik, Karl-Schwarzschild-Str. 1, 85748 Garching, Germany}
\affiliation{Astrophysics Research Institute, Liverpool John Moores University, 146 Brownlow Hill, Liverpool L3 5RF, UK}
\author{Maryam~Modjaz}
\affiliation{ Center for Cosmology and Particle Physics, Department of Physics, New York University, New York, NY 10003, USA}
\author[0000-0002-1092-6806]{Kishore C. Patra}
\affiliation{Department of Astronomy, University of California, Berkeley, CA 94720-3411, USA}
\affiliation{Nagaraj-Noll-Otellini Graduate Fellow}
\author[0000-0001-9710-4217]{Micalyn Rowe}
\affiliation{Mitchell Institute for Fundamental Physics and Astronomy,\\ Department of Physics and Astronomy, Texas A\&M University \\
4242 TAMU, College Station, TX, USA}
\author[0000-0001-5510-2424]{Nathan Smith}
\affiliation{Steward Observatory, University of Arizona, 933 North Cherry Avenue, Tucson, AZ 85721, USA} 
\author[0000-0001-9038-9950]{Schuyler~D.~Van~Dyk}
\affiliation{Caltech/IPAC, Mailcode 100-22, Pasadena, CA 91125, USA}
\author{Marc~Williamson}
\affiliation{ Center for Cosmology and Particle Physics, Department of Physics, New York University, New York, NY 10003, USA}
\author[0000-0002-6535-8500]{Yi Yang}
\affiliation{Department of Astronomy, University of California, Berkeley, CA 94720-3411, USA}
\affiliation{Bengier-Winslow-Robertson Postdoctoral Fellow}
\author{WeiKang~Zheng}
\affiliation{Department of Astronomy, University of California, Berkeley, CA 94720-3411, USA}
\author{Asia~deGraw}
\affiliation{Department of Astronomy, University of California, Berkeley, CA 94720-3411, USA}
\author[0000-0003-2238-1572]{Ori~D.~Fox}
\affiliation{Space Telescope Science Institute, 3700 San Martin Dr., Baltimore, MD 21218, USA}
\author[0000-0002-3739-0423]{Elinor L. Gates}
\affiliation{UCO/Lick Observatory, P.O. Box 85, Mount Hamilton, CA 95140, USA}
\author{Connor~Jennings}
\affiliation{Department of Astronomy, University of California, Berkeley, CA 94720-3411, USA},
\author[0000-0003-0427-8387]{R. Michael Rich}
\affiliation{Department of Physics and Astronomy, University of California, Los Angeles, CA 90095-1547, USA}

%\author{...}
%PLEASE ADD YOUR NAME HERE
\begin{abstract}
We present three epochs of early-time ultraviolet (UV) and optical {\it HST}/STIS spectroscopy of the young, nearby Type IIP supernova (SN) 2021yja. {We complement the {\it HST} data with two earlier epochs of {\it Swift} UVOT spectroscopy.} The {\it HST} and {\it Swift} UVOT spectra are consistent with those of other well-studied Type IIP supernovae (SNe). The UV spectra exhibit rapid cooling at early times, while less dramatic changes are seen in the optical. We also present Lick/KAIT optical photometry up to the late-time-tail phase, showing a very long plateau and shallow decline compared with other SNe~IIP. Our modeling of the UV spectrum with the \texttt{TARDIS} radiative-transfer code produces a good fit for a high-velocity explosion, a low total extinction $E(B-V) = 0.07$\,mag, and a subsolar metallicity. We do not find a significant contribution to the UV flux from an additional heating source, such as interaction with the circumstellar medium, consistent with the observed flat plateau. 
Furthermore, the velocity width of the \ion{Mg}{2} $\lambda$2798 line is comparable to that of the hydrogen Balmer lines, suggesting that the UV emission is confined to a region close to the photosphere. 
\end{abstract}


%%%%%%%%%%%%%%%%%%%%
%%  Introduction  %%
%%%%%%%%%%%%%%%%%%%%

\section{Introduction}\label{s:intro}
%-\textbf{Short review of CCSNe and their progenitors}\\
%Supernovae (SNe) are separated into two classes based on whether they have significant Hydrogen in their spectrum (Type II) or whether they do not (Type I). 
Type II supernovae (SNe~II) are defined by the presence of hydrogen in their spectra. They can be photometrically distinguished by their light-curve shape; an SN~II with a linearly (in magnitudes) declining light curve is designated as IIL, whereas an SN~II displaying an extended plateau lasting $\sim 90$\,days after explosion is classified as IIP. However, the  distinction between SNe~IIP and IIL is not clear, with recent works suggesting that these subtypes instead constitute a continuum \citep{anderson_characterizing_2014,valenti_diversity_2016}. The explosion mechanism producing Type II and a subset of Type I (Ib/Ic; stripped-envelope) SNe is widely accepted to be the core collapse of a star with zero-age main-sequence (ZAMS) mass $\geq 8$\,M$_{\odot}$; see \citet{filippenko_optical_1997} for a review. %\N{(note that the upper end of the mass range for RSGs+SNe II-P - and what happens above that - is not well known and controversial.  i.e. the red supergiant problem.  cite Smartt 2009, Davies \& Beasor, etc.)} %[cite: Woosley, S. E., & Weaver, T. A. 1986, ARA&A, 24, 205].
%On the other hand, Type Ia SNe  are thought to be produced by a runaway thermonuclear burning of a white dwarf. 
The spectral continuum of an SN~II peaks in the ultraviolet (UV) in the days and weeks following the explosion, and then continues to shift toward optical and near-infrared (NIR) wavelengths over the next few months through a combination of cooling and line blanketing.
%\indent

%\textbf{The following paragraph can be removed, this is a bit too basic for the paper. *Yi is suggesting that we keep this paragraph.* Alex, what do you think?}
%***\alex\Keep it.
It is generally accepted that the progenitors of SNe~IIP are red supergiants \citep[RSGs;][]{smartt_death_2009,dyk_supernova_2011, smartt_observational_2015,van_dyk_direct_2017}. 
%However, no SN~IIP progenitors have been found above an initial mass of $\sim 17$\,M$_{\odot}$, in what is known as the ``red supergiant problem" \citep{smartt_death_2009,davies_red_2020}. 
% this is not really true - 17 Msun is out of date, so ...
%\N{(suggested rewording of last sentence:)
However, among directly detected progenitors of SNe~IIP, progenitors with masses as high as the most massive RSGs seen in nearby stellar populations seem to be missing, in what is known as the ``red supergiant problem" \citep{smartt_death_2009,davies_red_2020}.
Preceding core collapse, an RSG maintains a significant fraction of its hydrogen envelope despite losing some of its initial mass to stellar winds. Upon collapse, the infalling material rebounds off of the newly formed neutron star and is further accelerated by interactions with neutrinos. A shock wave propagates outward, depositing 10--20\% of the neutrino energy (1--2 $\times 10^{52}$ ergs) %\N{(this is unclear - do you mean 10-20 per cent of the total neutrino energy?  it sounds like you are saying that 1e52 ergs is 10-20 per cent of the shock energy.) Sergiy: I think I clarified this now.} 
into the expansion of the material and leading to ejecta velocities that reach $\sim 0.1$c.
% and manifests as a blueshift in the spectrum's absorption features. 
A consequence of this explosion model is that one can approximate the expansion as homologous. Thus, assuming a spherically symmetric explosion, the radius of the photosphere (where optical depth $\tau = 2/3$) is proportional to the photospheric velocity, $r_{\rm ph} = v_{\rm ph}t$ \citep{kirshner_distances_1974,dessart_distance_2005}.

%\indent
Although optical spectra of core-collapse SNe (CCSNe) have been extensively studied, work on UV radiation has been relatively lacking. Such observations are challenging because rapid follow-up spectroscopy within $\sim 1$ week after the SN explosion is required from space-based telescopes, before the UV radiation peak has shifted to longer wavelengths. Additionally, SNe need to be sufficiently nearby to enable a decent signal-to-noise ratio (S/N) in the UV. However, several programs have made significant progress in UV spectroscopy of SNe~IIP, and we will refer to them throughout this work.

In general, CCSNe exhibit diverse UV/optical spectra. However, SNe~IIP have shown some uniformity in their UV spectra.  Works by the {\it Swift} and the {\it Galaxy Evolution Explorer (GALEX)} satellites have revealed similarities in the shapes of the UV spectra for SN~1999em-like SNe \citep[SNe~2005cs, 2005ay, 2006bp;]{bufano_ultraviolet_2009,gal-yam_galex_2008}. On the other hand, the sample is sparse and mostly comes from a subset of SNe observed by the {\it Hubble Space Telescope (HST)} and {\it Swift} UVOT programs.  SNe~IIL (e.g., SNe\,1979c and 1980K; \citealp{panagia_coordinated_1980}), although spectroscopically similar in the optical to SNe~IIP, have been shown to exhibit a UV excess below 1500~\AA, blueshifted \ion{Mg}{2} line emission, and a smooth continuum suggesting interaction with circumstellar material \citep[CSM;][]{panagia_coordinated_1980}.

 The UV spectra of SNe~IIP are not only similar in the shape of the continuum, but also in spectral features. This includes the prominent \ion{Mg}{2} $\lambda$2798 P~Cygni profile, as well as emission ``bumps" around 2200\,\AA, 2400\,\AA, and 2600\,\AA. These emission features are associated with blended \ion{Fe}{2} and \ion{Ni}{2} lines    \citep{brown_early_2007,gal-yam_galex_2008,bufano_ultraviolet_2009,dhungana_extensive_2016}.

 Early-time UV spectra convey important information about the kinematics of the fast-expanding ejecta, the temporal evolution of the photospheric temperature, and the metal content of the progenitor star \citep{mazzali_applications_2000,dessart_quantitative_2005,dessart_quantitative_2006}. 
 %\N{(say someting about why the early temperature and fast expanding ejecta are importnat... i.e. what do they tell you about the progenitor;s envelope or the shock breakout...or?)} 
 %***\alex\Could deal with some of Nathan's comments during the refereeing stage (after receiving the report). Not necessarily now.
 The line-of-sight extinction may also be well determined through spectroscopic modeling of the UV. Moreover, the UV flux is an excellent probe of the circumstellar environment in the vicinity of the SN, allowing one to identify additional heating sources such as CSM interaction \citep{ben-ami_ultraviolet_2015}. In this work, we
 %***\alex\We have Swift UV spectra, too, so I changed 3 to 5.
 present five early-time UV spectra along with optical spectra and photometry of a relatively nearby SN~IIP.

SN~2021yja (AT~2021yja) was discovered on 8 Sep. 2021 at 13:12:00 (UTC dates are used throughout this paper) in the spiral galaxy NGC~1325 by the Asteroid Terrestrial-impact Last Alert System \citep[ATLAS;]{tonry_atlas_2018}; see AstroNote 2021--235 \citep{smith_atlas21bidw_2021}. A redshift of $z = 0.005307$ was reported by \citet{springob_digital_2005}, and a median distance of 21.8\,Mpc can be queried from the NASA/IPAC NED Database.  Imaging obtained with the  MuSCAT3 instrument on the Faulkes-North Telescope (FTN) at Haleakala, Hawai`i, reported a transient consistent with SN~2021yja on 7 Sep. 2021 at 15:02:28 (AstroNote 2021-236; \citealt{kilpatrick_at2021yja_2021}). We adopt the midpoint between the last nondetection (6 Sep. 2021 at 11:32:38) and the first detection (8 Sep. 2021 at 13:12:00) as an estimated time of explosion, 7.5 Sep. 2021. All phases will be given in days relative to this date throughout the paper.\

\begin{figure}
    \centering
    \includegraphics[width=0.5\textwidth]{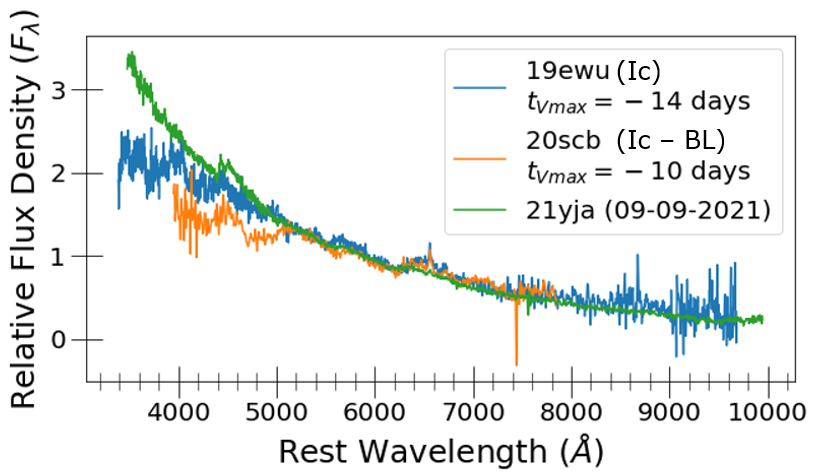}
    \caption{Very-early-time spectroscopy of SN~2021yja (green) compared to Type Ic/Ic-BL SNe 2019ewu (blue; \citealt{hiramatsu_global_2019}; {Williamson et al., in prep.}) and 2020scb (orange; \citealt{smith_atlas20yki_2020}).}
    \label{fig:tns_comp}
\end{figure}

In Figure \ref{fig:tns_comp}, we compare our earliest optical spectrum of SN~2021yja obtained on 9.5 Sep. 2021 (+2\,days) by the FLOYDS spectrograph mounted on the FTN \citep{pellegrino_global_2021} with early-time spectra of Type Ic SN~2019ewu (\citet{hiramatsu_global_2019}; {Williamson et al., in prep.}) and Type Ic-BL SN~2020scb \citep{smith_atlas20yki_2020} at similar phases; there is a close resemblance.  Although weak, broad H$\alpha$ emission might be present in the spectrum of SN~2021yja, this feature is even more prominent in the spectra of SNe~2019ewu (Ic) and SN~2020scb (Ic-BL). 

Moreover, all three objects exhibit a small bump near 4600~\AA\ superposed on a blue, otherwise relatively featureless continuum.
Hence, we triggered SN~2021yja for our Cycle 28 {\it HST} program that targeted stripped-envelope SNe (GO-16178; PI A. V. Filippenko). 

Later, it became clear that the optical spectrum of SN~2021yja was transforming into that of an SN~II, and the optical photometry was consistent with the Type IIP subtype,
but the {\it HST} observations had already begun and the object was deemed sufficiently interesting to continue monitoring.
% We show below that the object fits the Type IIP subtype from our photometric analysis.
%however, the early-time optical spectrum was not clearly distinguishable from that of some SNe~Ib/Ic. 
The spectra presented in this paper make SN~2021yja one of the few SNe~IIP studied with early (9\,days after explosion), high-S/N {($\sim 35$)}, UV data.  

The paper is organised as follows. Section 2 presents further discussion of this topic and a summary of our observations. In Section 3, we discuss the spectroscopic analysis of SN~2021yja using a Monte-Carlo radiative transfer code, \texttt{TARDIS}. We conclude in Section 4.

%%%%%%%%%%%%%%%%%%%%
%%  Summary of Observations and Data Analysis  %%
%%%%%%%%%%%%%%%%%%%%
% \newpage
\section{Observations and Data Analysis}\label{sec:obs_analysis}

\subsection{HST/STIS}
\label{sec:hst_stis}
%In this section, we present data from our {\it HST} program.
SN~2021yja was observed as an {\it HST} disruptive target of opportunity on 16 Sep. 2021 ({+9}\,days after explosion), 21 Sep. 2021 (+14\,days), and 28 Sep. 2021 (+21\,days) using the CCD ($52^{\prime \prime} \times 52^{\prime \prime}$ field of view) detector of the Space Telescope Imaging Spectrograph (STIS). Although the onboard Near-UV Multi-Anode MicroChannel Array (NUV-MAMA) detectors do not suffer from read noise and charge-transfer efficiency (CTE) like the CCDs, they do have bright-object limits \citep{stis}.
We used the CCD mode to avoid saturating the detector (although, in retrospect, the UV flux had fallen sufficiently by the epoch of the first {\it HST} observation), and also because there were fewer scheduling constraints (we did not have to avoid the South Atlantic Anomaly).

For Epochs 1 and 3, observations of the mid-UV (MUV ; 1685--3060\,\AA) with the G230LB grating were made over 6 visits, whereas Epoch 2 had only 5 visits. Only one visit per epoch was made for the Near-UV (NUV ; 2900--5700\,\AA ) and optical (5240--10270\,\AA) with the G430L and G750L gratings, respectively. A detailed observation log can be found in Table \ref{Table:HSTobslog}, and the three {\it HST} spectra are shown in Figure~2.\\

\begin{deluxetable}{lccc}
\label{Table:HSTobslog}
    %\tablewidth{0.9\textwidth}
    \tablecaption{HST Observation Log for SN~2021yja}
    \tablehead{ \colhead{Date [UTC]} & \colhead{Exp. [s]}  & \colhead{Grating/Filter} & \colhead{$\Delta \lambda$ [\AA]}  }
    \startdata
        2021-09-16 & 2060 & G230LB & 1685-3060 \\
        2021-09-16 & 2340 & G230LB & 1685-3060\\
        2021-09-16 & 2340 & G230LB & 1685-3060\\
        2021-09-16 & 2000 & G230LB & 1685-3060\\
        2021-09-16 & 2340 & G230LB & 1685-3060\\
        2021-09-16 & 1544 & G230LB & 1685-3060\\
        2021-09-16 & 200 & G430L & 2900-5700\\
        2021-09-16 & 100 & G750L & 5240-10,270\\
        \hline
        2021-09-21 & 2060 & G230LB & 1685-3060\\
        2021-09-21 & 2340 & G230LB & 1685-3060\\
        2021-09-21 & 2340 & G230LB & 1685-3060\\
        2021-09-21 & 2000 & G230LB & 1685-3060\\
        2021-09-21 & 1544 & G230LB & 1685-3060\\
        2021-09-21 & 200 & G430L & 2900-5700\\
        2021-09-21 & 100 & G750L & 5240-10,270\\
        \hline
        2021-09-28 & 2060 & G230LB & 1685-3060\\
        2021-09-28 & 2340 & G230LB & 1685-3060\\
        2021-09-28 & 2340 & G230LB & 1685-3060\\
        2021-09-28 & 2000 & G230LB & 1685-3060\\
        2021-09-28 & 2340 & G230LB & 1685-3060\\
        2021-09-28 & 1544 & G230LB & 1685-3060\\
        2021-09-28 & 200 & G430L & 2900-5700\\
        2021-09-28 & 100 & G750L & 5240-10,270\\
    \enddata
    % \label{Table:HSTobslog}
    %\tablecomments{HST/STIS Observation log of 2021yja for the three epochs.}
\end{deluxetable}

%***\alex\In Figure 1, insert label, change Ic-bl to Ic-BL (the usual convention, I think)... can be done after submission. 

%***\alex\
%Put "f_\lambda" after "Flux Density" in the ordinate label of Figure 2, just to make it very clear.
% Put a space between 9100 and km/s in the label
\begin{figure*}
    \centering
    \includegraphics[width=1.0\textwidth]{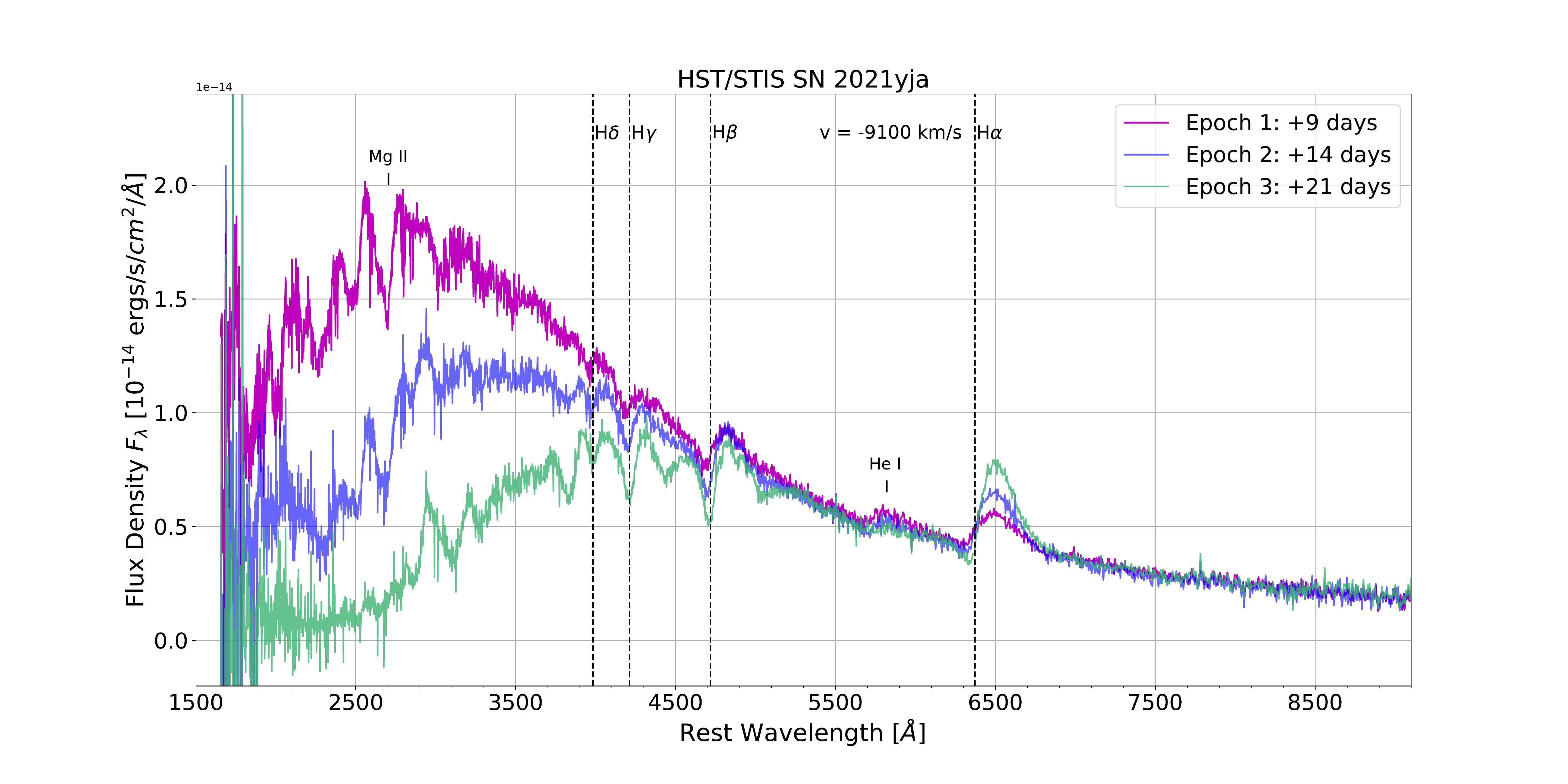}
    \caption{{\it HST}/STIS UV-optical spectra of SN~2021yja. No normalization has been applied; in other words, there is general agreement in flux density at optical wavelengths while the UV varies dramatically. The wavelength scale has been corrected to the rest frame using the recession velocity of the host galaxy. Balmer lines at an expansion velocity of $v = -9100$\,km\,s$^{-1}$ are marked by vertical dotted lines.}
    \label{fig:hst_uvo}
\end{figure*}

\newpage
\subsection{Lick Observatory Kast Spectra and KAIT Photometry}
\label{sec:kastkait}

\begin{figure*}[h]
    \centering
    \includegraphics[width=0.9\textwidth]{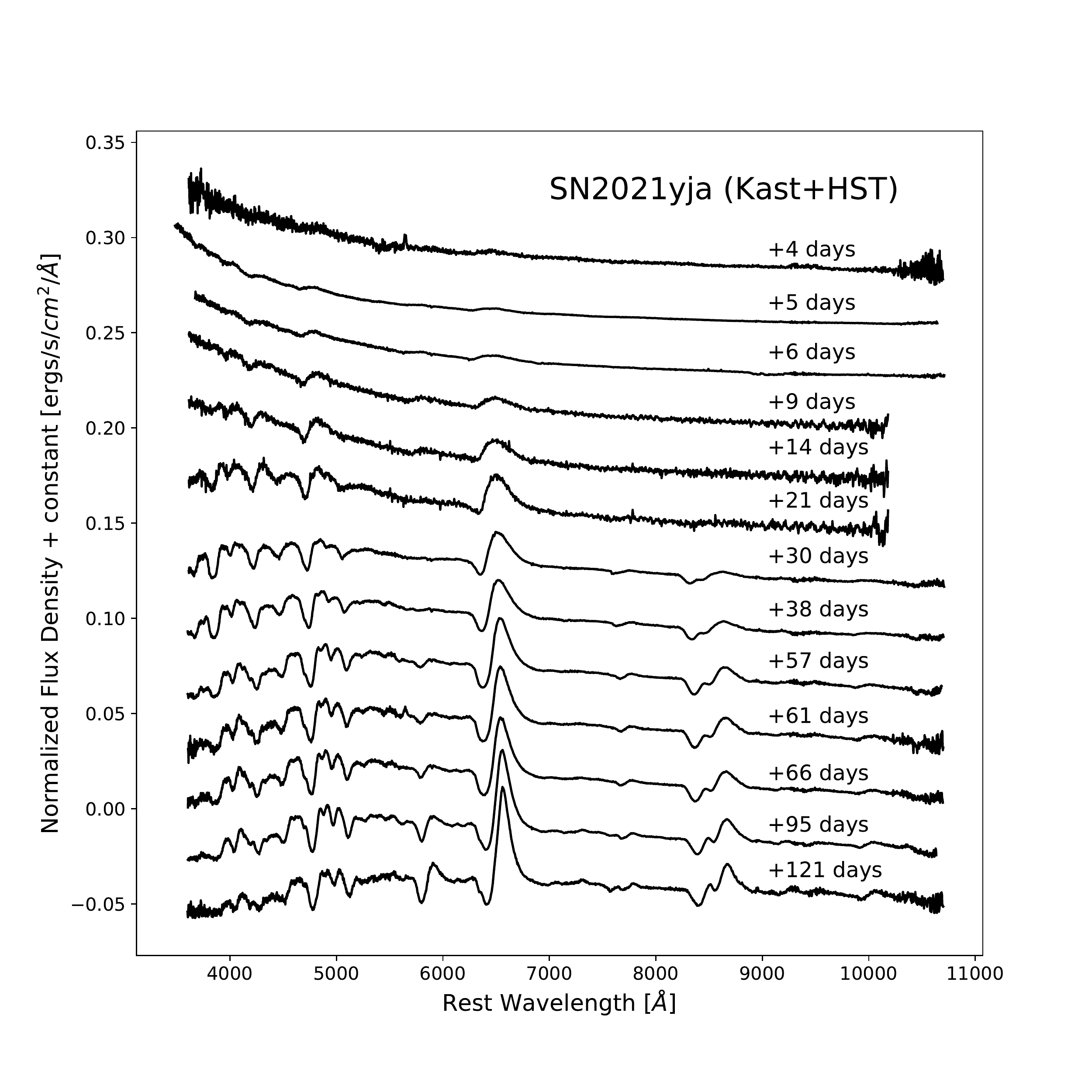}
    \caption{Optical spectral time series of SN~2021yja observed with Kast and {\it HST}/STIS (days +9, +14, and +21). The wavelength scale has been corrected to the rest frame using the recession velocity of the host galaxy.}
    \label{fig:kast_hst}

\end{figure*}

Optical spectra of SN~2021yja were obtained using the Kast Double Spectrograph on the Shane 3\,m telescope at Lick Observatory \citep{miller_ccd_1988, miller_93}. The spectral sequence consists of 10 epochs and spans +4 to +121\,days. All observations cover a wavelength range from 3632 to 10,754\,\AA. Observations and data reduction were carried out following the techniques described by \citet{silverman_berkeley_2012} and \citet{shivvers_nebular_2013}. The Kast observation log is shown in Table \ref{tbl:specpol_log}. To facilitate cosmic-ray removal, observations consisted of three exposures in the red channel (which employs a thick CCD), each a third of the exposure time in Table \ref{tbl:specpol_log} (but note that only a single red-channel exposure was obtained on 13 Sep. 2021). The blue channel (thin CCD) was exposed throughout the red-channel observations, with an additional 60\,s to synchronize readout times. The phase is rounded to the nearest day relative to explosion. The optical spectra are shown in Figure \ref{fig:kast_hst}. %\N{(you mean Fig 3, right? Yes, whoops)}

\begin{table*}
\caption{Kast Spectroscopy of SN~2021yja.$^a$} 
\begin{center}

\begin{tabular}{cccccccc}
	\hline 
	\hline
	UT Date & Phase$^b$ & Average & Seeing & Red-Chan. Total  \\ 
	(MM-DD-YYYY)& (days) & Airmass  & (arcsec) & Exposure (s)$^c$ \\ 
	\hline 
    09-11-2021 & 4 & 1.9 & 3 &  900 \\
    09-12-2021 & 5 & 1.9 & 1.7  & 900     \\
    09-13-2021 & 6 & 2.0 & 2 & 600$^d$ \\
    10-07-2021 & 30 & 2.0 &  2 & 500   \\
    10-15-2021 & 38 & 2.0  & 2 & 1200  \\
    11-03-2021 & 57 &  2.4 & 1.2  &  1200  \\
    11-07-2021 & 61 & 3.0  & 2.6   & 1200  \\
    11-12-2021 & 66 & 2.2  &  1.2  &  1200 \\
    12-11-2021 & 95 & 2.0  & 1.3  &   1200\\
    01-06-2022 & 121 &  & 1.2  &  1500\\
    
	\hline 
\end{tabular}\\
\end{center}
%\flushleft
\centering
{$^a$}{The wavelength range was 3632--10,754\,\AA\ for each observation.} \\
{$^b$}{Days after explosion assuming  09-07-2021 as the explosion date.} \\
{$^c$}{The blue-channel exposure time was 60\,s longer on each date.}\\
{$^d$}{Only one exposure was taken with the red channel.}

\label{tbl:specpol_log}
\end{table*}

Follow-up photometry of SN~2021yja was performed with images from the 0.76\,m Katzman Automatic Imaging Telescope
(KAIT) as part of the Lick Observatory Supernova Search \citep[LOSS;][]{filippenko_lick_2001}, as well as with images from the 1\,m Nickel telescope at Lick Observatory.
$B$, $V$, $R$, and $I$ multiband images were obtained with both telescopes, and
additional {\it Clear}-band images \citep[similar to $R$;][]{li_nearby_2011}
were obtained with KAIT.

\begin{figure*}
    \centering
    \includegraphics[width=\textwidth]{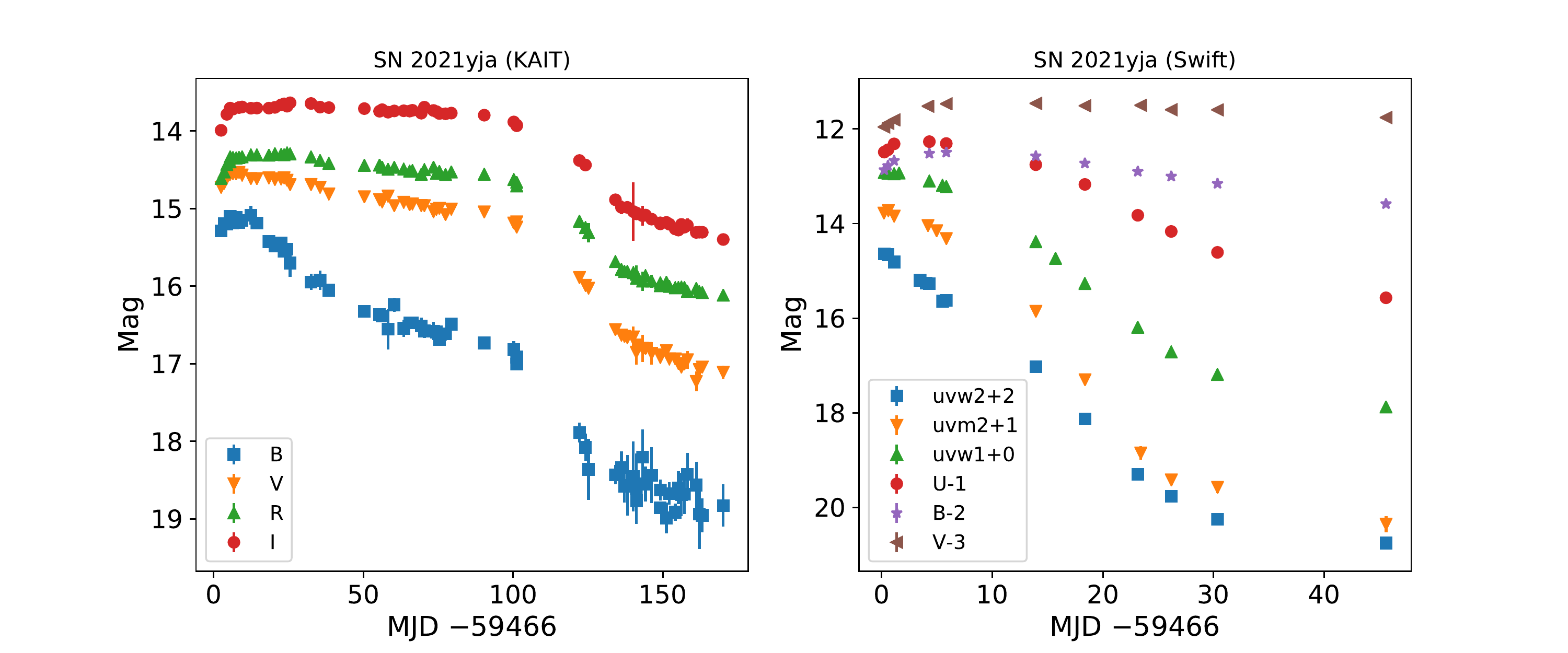}
    \caption{KAIT and {\it Swift} photometry of SN~2021yja in apparent magnitudes. No extinction corrections have been applied to the data.}
    \label{fig:kait_swift_phot}
\end{figure*}

All images were reduced using a custom
pipeline\footnote{https://github.com/benstahl92/LOSSPhotPypeline}
detailed by \citet{stahl_lick_2019}.
%Here we briefly summarize the photometry procedures:
%Image subtraction procedure was applied in order to remove the host-galaxy light,
%using additional images obtained after the SN has faded below our detection limit.
Point-spread-function (PSF) photometry was obtained using DAOPHOT \citep{stetson_daophot_1987}
from the IDL Astronomy User's Library\footnote{http://idlastro.gsfc.nasa.gov/}.
Several nearby stars were chosen from the
Pan-STARRS1\footnote{http://archive.stsci.edu/panstarrs/search.php} catalog for calibration.
Their magnitudes were first transformed into Landolt magnitudes
\citep{landolt_ubvri_1992}
using the empirical prescription presented by \citet[][see Eq. 6]{tonry_pan-starrs1_2012},
and then transformed to the KAIT/Nickel natural system.
Apparent magnitudes were all measured in the KAIT4/Nickel2 natural system.
The final results were transformed to the standard system using local
calibrators and color terms for KAIT4 and Nickel2 \citep{stahl_lick_2019}. The optical light curves are presented in the left panel of Figure \ref{fig:kait_swift_phot}.


\begin{deluxetable}{lccccc}
    %\tablewidt<h{1.0\textwidth}
    \tabletypesize{\scriptsize}
    %\tablewidth{0pt}
    \tablecaption{KAIT/Nickel Photometry of SN~2021yja}
    \tablehead{ \colhead{Date} & \colhead{$B$}  & \colhead{$V$} & \colhead{$R$} & \colhead{$Clear$} & \colhead{$I$} \\
    \colhead{[MJD]} & \colhead{[mag]}  & \colhead{[mag]} & \colhead{[mag]} & \colhead{[mag]} & \colhead{[mag]}}
    \startdata
        59468.51 & 14.79(03) &	14.72(01) &	14.61(01) &	14.47(01) &	14.49(01)\\
        59469.54 & 14.69(04) &	14.67(02) &	14.53(02) &	14.41(05) &	   ...\\
        59470.44 & 14.70(02) &	14.60(01) &	14.43(01) &	14.39(02) &	14.29(01)\\
        59471.49 &	14.69(02) &	14.56(01) &	14.40(01) &	   ...	  & 14.22(02)\\
        59471.54$^{*}$ &	14.60(02) &	14.48(01) &	14.34(01) &	   ...    & 14.21(01) \\       
        59472.46 &	14.69(06) &	14.55(01) &	14.35(01) &	14.29(02) &	14.22(01)\\
        59473.53 &	14.64(04) &	14.55(01) &	14.35(01) &	14.32(03) &	   ...\\
        59474.50 &	14.68(03) &	14.55(02) &	14.34(01) &	14.25(01) &	14.20(01)\\
        59475.53 &	14.65(03) &	14.57(01) &	14.34(01) &	14.28(02) &	14.19(02)\\
        59478.47 &	14.59(12) &	14.61(03) &	14.31(02) &	14.30(02) &	14.21(02)\\
        59480.45 &	14.69(05) &	14.61(02) &	14.31(01) &	14.27(03) &	14.21(02)\\
        59484.47 &	14.93(03) &	14.60(02) &	14.31(01) &	14.31(02) &	14.21(02)\\
        59486.48 &	14.98(04) &	14.62(01) &	14.30(01) &	14.29(01) &	14.20(02)\\
        59488.55 &	14.95(05) &	14.62(02) &	14.31(01) &	14.34(01) &	14.16(02)\\
        59489.54 &	15.05(05) &	14.60(02) &	14.31(01) &	14.32(01) &	14.15(02)\\
        59490.54 &	15.02(04) &	14.63(02) &	14.29(01) &	14.34(01) &	14.18(02)\\
        59491.55 &	15.20(18) &	14.69(06) &	   ...    &    ...	  & 14.14(03)\\
        59498.54 &	15.45(10) &	14.69(03) &	14.34(03) &	14.41(02) &	14.15(04)\\
        59501.55 &	15.42(13) &	14.72(03) &	14.38(03) &	14.46(21) &	14.19(03)\\
        59504.49 &	15.55(04) &	14.81(02) &	14.42(01) &	14.45(01) &	14.20(02)\\
        59516.34	&	15.83(03) &	14.85(01) &	14.45(01) &	14.49(01) &	14.21(02)\\
        59521.41	&	15.86(04) &	14.88(02) &	14.44(02) &    ...    & 14.25(02)\\
        59522.29	&	15.88(04) &	14.91(02) &	14.47(01) &	14.52(01) &	14.23(02)\\
        59524.24	&	16.06(26) &	14.84(05) &	14.50(03) &	14.52(02) &	14.26(03)\\
        59526.32	&	15.74(09) &	14.96(03) &	14.47(02) &	...       &	14.24(02)\\
        59529.48	&	16.04(11) &	14.92(04) &	14.49(02) &	14.59(02) &	14.24(03)\\
        59531.45	&	15.97(04) &	14.95(02) &	14.52(01) &	14.56(01) &	14.24(01)\\
        59532.39	&	15.97(05) &	14.94(02) &	14.51(01) &	14.60(03) &	14.23(02)\\
        59535.34	&	16.01(11) &	14.96(04) &	14.56(03) &	14.60(03) &	14.27(04)\\
        59536.37	&	16.08(09) &	14.96(03) &	14.50(02) &	14.63(03) &	14.19(03)\\
        59539.41	&	16.08(12) &	15.04(04) &	14.47(04) &	14.57(05) &	14.24(05)\\
        59540.41	&	16.09(09) &	15.01(03) &	14.55(02) &	14.57(02) &	14.25(03)\\
        59541.37	&	16.19(05) &	15.00(02) &	14.52(02) &	14.59(01) &	14.28(02)\\
        59543.44	&	16.12(08) &	15.08(03) &	14.56(02) &	14.60(02) &	14.28(02)\\
        59545.36	&	15.99(07) &	15.01(02) &	14.53(02) &	14.58(01) &	14.27(02)\\
        59556.36	&	16.23(05) &	15.04(02) &	14.56(02) & 14.65(02) &	14.30(02)\\
        59566.22	&	16.32(11) &	15.19(05) &	14.63(06) &	14.73(05) &	14.38(05)\\
        59567.20	&	16.41(09) &	15.24(02) &	14.66(02) &	14.77(01) &	14.43(02)\\
        59588.14	&	17.39(13) &	15.89(04) &	15.17(02) &	15.30(01) &	14.88(02)\\
        59590.15	&	17.58(18) &	15.99(04) &	15.25(03) &	15.40(03) &	14.94(03) \\    
        59600.15	&	17.93(12) &	16.56(03) &	15.69(02) &	15.87(14) &	15.39(03)\\
        59602.14	&	17.84(21) &	16.63(05) &	15.79(05) &	15.90(02) &	15.48(09)\\
        59603.17	&	18.08(21) &	16.65(05) &	15.81(03) &	15.92(13) &	...\\
        59604.12	&	18.07(39) &	16.66(05) &	15.81(02) &	15.94(01) &	15.49(03)\\
        59606.14	&	17.95(45) &	16.65(13) &	15.83(04) &	16.02(03) &	15.54(37)\\
        59607.11	&	18.11(46) &	16.85(16) &	15.85(12) &	16.01(02) &	15.56(04)\\
        59607.19$^*$	&	18.27(04) &	16.76(02) &	15.90(02) &	  ...     &	15.57(02)\\
        59609.21	&	17.70(36) &	16.80(18) &	15.94(13) &	16.04(02) &	15.59(13)\\
        59610.19	&	18.05(22) &	16.80(08) &	15.87(03) &	16.04(02) &	15.58(03)\\       
        59612.23	&	17.94(36) &	16.87(15) &	15.94(08) &	16.05(08) &	15.64(07)\\
        59615.11$^*$	&	18.36(05) &	16.88(02) &	16.00(02) &	 ...  &	15.68(02)\\
        59615.14	&	18.13(14) &	16.92(04) &	15.96(02) &	16.11(05) &	15.70(04)\\
        59617.17	&	18.49(20) &	16.83(05) &	15.95(03) &	16.13(02) &	15.68(03)\\
        59618.15	&	18.17(14) &	16.94(04) &	16.01(03) &	16.15(04) &	15.70(03)\\
        59620.11$^*$	&	18.41(11) &	16.93(03) &	16.02(03) &	  ...     &	15.76(03)\\
        59621.16	&	18.10(21) &	17.00(07) &	16.02(04) &	16.20(07) &	15.78(06)\\
        59622.17	&	18.19(28) &	17.04(06) &	16.02(03) &	16.19(08) &	15.71(05)\\
        59623.17	&	18.18(25) &	17.00(06) &	16.02(03) &	16.19(08) &	15.74(04)\\
        59624.17	&	17.93(27) &	16.95(12) &	16.07(06) &	16.18(09) &	15.72(09)\\
        59627.13	&	18.07(30) &	17.23(13) &	16.03(05) &	16.27(04) &	15.81(05)\\
        59628.14	&	18.44(45) &	17.07(08) &	16.07(03) &	16.27(03) &	15.81(04)\\
        59629.14	&	18.44(22) &	17.04(06) &	16.09(03) &	16.26(02) &	15.81(03)\\
    \enddata
    \tablecomments{$^*$Observed with Nickel.}% Continued in \textbf{Table} \ref{Table:kaitphot2}.}
    \label{Table:kaitphot}
\end{deluxetable}

\subsection{Swift UVOT}
\begin{deluxetable*}{lcccccc}
    %\tablewidth{0.9\textwidth}
    
    \tablecaption{{\it Swift} UVOT Photometry of SN~2021yja}
    \tablehead{ \colhead{Date [MJD]} & \colhead{$uvw2$ [mag]}  & \colhead{$uvm2$ [mag]} & \colhead{$uvw1$ [mag]} & \colhead{$U$ [mag]} & \colhead{$B$ [mag]} & \colhead{$V$ [mag]} }
    \startdata
        59466.2 & 12.64(04) & 12.78(04) & 12.93(04) & 13.49(04) & 14.87(05) & 14.96(06)\\
        59466.6 & 12.66(04) & 12.72(05) & 12.94(04) & 13.44(04) & 14.78(05) & 14.88(06)\\
        59467.1 & 12.81(04) & 12.84(04) & 12.95(04) & 13.32(04) & 14.67(04) & 14.81(06)\\
        59467.6 & ... & ... & 12.94(04) & ... & ... & ...\\
        59469.5 & 13.20(04) & ... & ... & ... & ... & ...\\
        59470.0 & 13.26(04) & ... & ... & ... & ... & ...\\
        59470.2 & ... & 13.04(04) & ... & ... & ... & 14.52(06)\\
        59470.3 & 13.27(04) & ... & 13.10(04) & 13.27(04) & 14.52(04) & ...\\
        59471.0 & ... & 13.15(05) & ... & ... & ... & ...\\
        59471.5 & 13.64(04) & ... & 13.19(04) & ... & ... & ...\\
        59471.8 & 13.63(04) & 13.32(05) & 13.22(04) & 13.31(04) & 14.50(04) & 14.47(06)\\
        59479.9 & 15.03(06) & 14.85(06) & 14.39(06) & 13.76(04) & 14.57(04) & 14.46(06)\\
        59484.4 & 16.13(07) & 16.30(07) & 15.27(06) & 14.17(05) & 14.73(04) & 14.51(06)\\
        59489.2 & 17.30(08) & ... & 16.19(06) & 14.82(05) & 14.90(05) & ...\\
        59489.4 & ... & 17.85(14) & ... & ... & ... & 14.50(06)\\
        59492.2 & 17.77(09) & 18.42(12) & 16.71(08) & 15.17(06) & 15.00(05) & 14.59(06)\\
        59496.4 & 18.25(11) & 18.57(12) & 17.19(09) & 15.61(07) & 15.16(05) & 14.60(06)\\
        59511.6 & 18.75(13) & 19.35(17) & 17.88(12) & 16.57(08) & 15.58(05) & 14.76(05)\\
    \enddata
    \tablecomments{{\it Swift}/UVOT Photometry of 2021yja as presented by \citet{hosseinzadeh_weak_2022}. Apparent magnitudes are reported with their $1\sigma$ uncertainties as indicated by the number in parentheses and in units of 0.01.}
    \label{Table:swiftphot}
\end{deluxetable*}

% SWIFT OBSERVATION AND DATA ANALYSIS

SN~2021yja was observed photometrically and spectroscopically in the UV with the {\it Neil Gehrels Swift Observatory} \citep{gehrels_theswiftgamma-ray_2004}. The Ultraviolet/Optical Telescope (UVOT; \citealp{roming_swift_2005}) is a Ritchey-Chr\'{e}tien reflector with a CCD detector. Its wavelength range covers 1600--6000\,\AA\ with imaging capabilities in six bandpasses ($uvw2$, $uvm2$, $uvw1$, $u$, $b$, and $v$).
{\it Swift} observations began on 9 Sep. 2021 at 04:42:34.  The UVOT photometry was reduced using the pipeline of the {\it Swift} Optical Ultraviolet Supernova Archive \citep[SOUSA;][]{brown_sousa_2014} with the updated sensitivity corrections and an aperture correction calculated using observations from 2021. {\it Swift} photometry is presented in Figure \ref{fig:kait_swift_phot}, and also by {\citet{hosseinzadeh_weak_2022}}.

\begin{figure}
    \centering
    \includegraphics[width=0.5\textwidth]{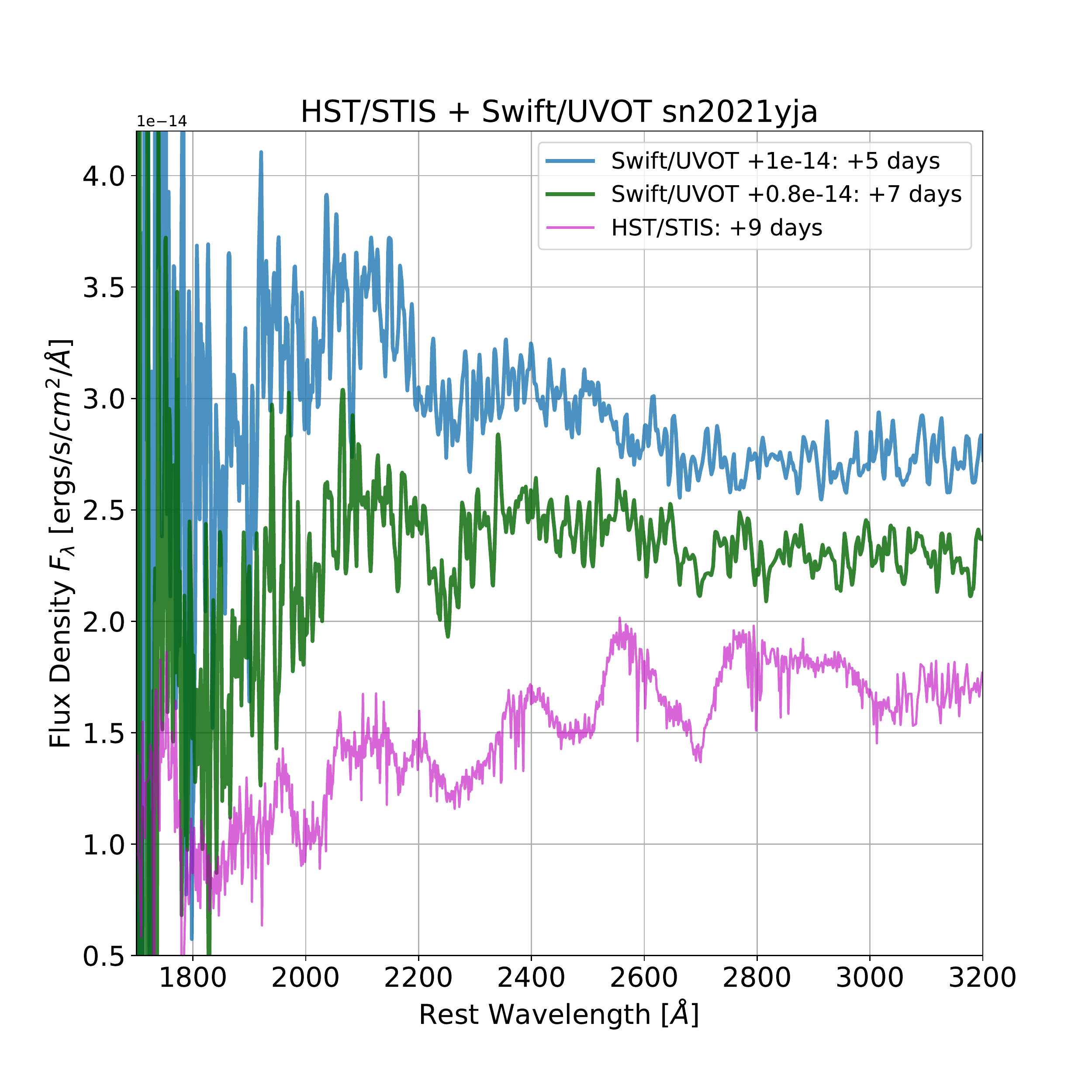}
    \caption{{\it HST}/STIS NUV spectrum of SN~2021yja at +9\,day {(S/N $\approx 35$)} and the median-combined {\it Swift}/UVOT spectra of SN~2021yja at +5 and +7\,days {(S/N $\approx 3$)} as described in the text. {\it Swift}/UVOT spectra were arbitrarily shifted for clarity.}
    \label{fig:hst_swift_uv}
\end{figure}

{\it Swift}/UVOT spectroscopy began on 12 Sep. 2021 1:10:19. Data were reduced using the \texttt{uvotpy} package \citep{kuin_uvotpy_2014} and the calibration of \citet{kuin_calibration_2015}.
Table~\ref{table:swiftgrism} summarizes the {\it Swift}/UVOT spectroscopy of SN~2021yja. Spectra collected within $\sim 20$\,hr were median combined to increase the S/N {to $\sim 3$ for wavelengths 1800--3600\,\AA}. A total of two spectra (at +5 and +7\,days) were obtained and used in further analysis (see Fig.~\ref{fig:hst_swift_uv}) .

\begin{deluxetable}{cccc}
%\tabletypesize{\scriptsize}
\tablecaption{{\it Swift} UVOT Spectroscopic Observations\label{table:swiftgrism}}
\tablehead{\colhead{Start Time} & \colhead{Obs ID} & \colhead{Exp.} & \colhead{Phase} \\ 
\colhead{(mm-dd-yyyy hh:mm:ss)} & \colhead{} & \colhead{(s)} & \colhead{(days)} } 

\startdata
09-12-2021 1:10:19 & 14808007 & 999.7 & 4.5 \\
09-12-2021 15:17:18 & 14808007 & 1322.6 & 5 \\
09-12-2021 17:04:22 & 14808007 & 709.4 & 5 \\
09-12-2021 21:46:21 & 14808007 & 978.9 & 5 \\
\hline
09-13-2021 23:22:18 & 14808007 & 946.6 & 6.5 \\
\hline
09-14-2021 10:17:18 & 14808010 & 1253.7 & 7 \\
09-14-2021 12:05:18 & 14808010 & 930.9 & 7 \\
09-14-2021 13:42:19 & 14808010 & 930.9 & 7 \\
\enddata

\tablecomments{Spectra obtained on 09-12-2021 and 09-14-2021 were median combined to increase the S/N, yielding spectra at +5 and +7\,days, respectively. The summed spectra are used in Figure 5.}

%% No \tablerefs indicated

\end{deluxetable}
% Swift grism observations
%SN2021yja_20210912_01.dat

%SN2021yja_20210912_15.dat
%SN2021yja_20210912_17.dat
%SN2021yja_20210912_21.dat

%SN2021yja_20210912_sum.dat  sum of the above 3

%SN2021yja_20210913_23.dat

%SN2021yja_20210914_10.dat
%SN2021yja_20210914_12.dat
%SN2021yja_20210914_13.dat

%SN2021yja_20210914_sum.dat  sum of the above 3
%%%% It looks like the two summed images are used in the paper.  

\begin{figure*}
    \centering
    \includegraphics[width=1.0\textwidth]{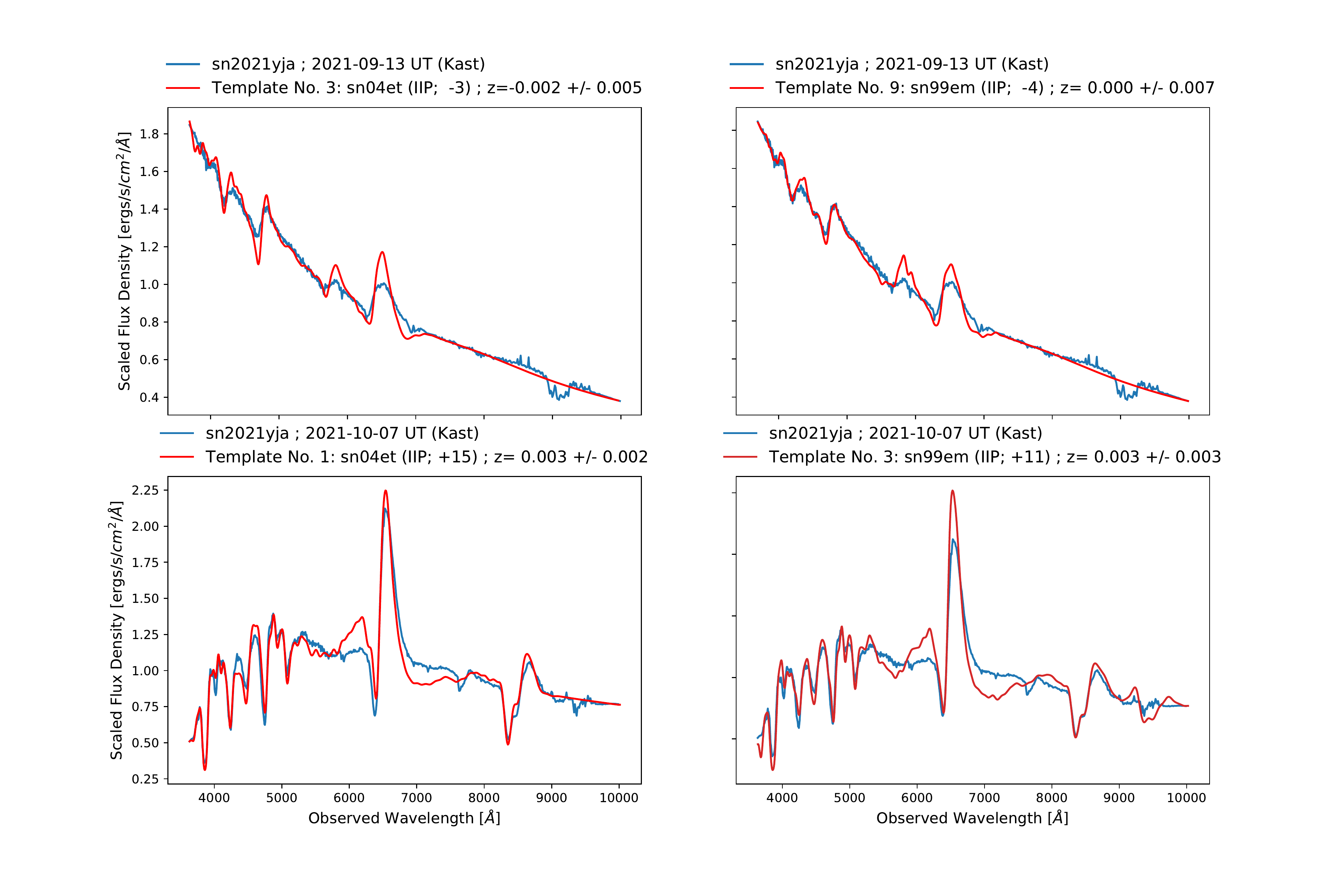}
    \caption{\textsc{SNID} best-fit templates for SN~2021yja on 13 Sep. 2021 (top), 7 Oct. 2021 (bottom).}
    \label{fig:snid}
\end{figure*}

\section{Photometric and Spectroscopic Analysis}

We ran the Supernova Identification \citep[SNID;][]{blondin_determining_2007} program on our observed Kast spectra of SN~2021yja. Cross-correlation with a library of SN spectra shows that the spectra of SN~2021yja obtained after +6\,days exhibit best matches with the Type IIP SN templates at similar phases (e.g., see Fig.~\ref{fig:snid}). The matching of SN~2021yja at even earlier phases failed to converge since the spectra are almost featureless in the optical. Throughout the rest of the paper, we classify SN~2021yja as an SN~IIP, as corroborated by the photometry presented in this section.

Galactic reddening toward SN~2021yja is $E(B-V)_{\text{MW}} = 0.02$\,mag \citep{schlafly_measuring_2011}. We find a small contribution to the extinction from the host galaxy ($E(B-V)_{\text{host}} = 0.05$\,mag), such that $E(B-V)_{\text{tot}} = 0.07$\,mag (see Sec. \ref{s:tardis}).  Spectra are dereddened unless specified otherwise. All of the
spectra are also corrected for the recession of the host galaxy NGC~1325 using
$z = 0.0053$ (NED/IPAC Extragalactic Database23\footnote{See \url{https://ned.ipac.caltech.edu/}}; \citealt{smith_atlas21bidw_2021}) unless specified otherwise. A detailed analysis of the optical light curves and spectra is presented by {\citet{hosseinzadeh_weak_2022}}.

\subsection{Photometric Comparison with Other SNe}
% In this section, we examine the light curves of SN~2021yja through a comparison with those of other well-sampled Type II and Type IIP SNe. 
In Figure \ref{fig:lccomp}, we compare the $V$-band light curve of SN~2021yja with those of the {canonical} Type IIL SN\,1979C \citep{vaucouleurs_bright_1981} and the Type IIP SNe 1999em \citep{leonard_distance_2002}, 2004et \citep{maguire_optical_2010}, 2005cs \citep{pastorello_sn_2009}, 2012aw \citep{valenti_supernova_2015}, 2013ej \citep{huang_sn_2015}, and 2017eaw \citep{dyk_type_2019}. {The comparison SNe are well-studied and occupy a wide range of light-curve parameters.} We apply an extinction correction to each SN. Several studies based on large samples of SN light curves have found that the duration of the $V$-band plateau is shorter for those SNe~IIP that exhibit higher $V$-band peak luminosity \citep{anderson_characterizing_2014,valenti_diversity_2016,dejaeger_berkeley_2019}. SN~2021yja displays a long plateau phase characteristic of SNe~IIP; however, the plateau dropoff is shallower compared to those of the other SNe. The absolute $V$ magnitude, $M_V$, peaks at nearly $-17.5$ mag, resembling the average behavior of normal SNe~IIP. We note that the object's distance (and therefore derived quantities such as absolute magnitude) are somewhat uncertain, since it does not have a Cepheid or tip-of-the-red-giant-branch distance estimate to its host.

The absolute $V$ magnitude at the start of the radioactive tail phase, $M_{\text{tail}}\approx -15.5$, is higher than what is measured in other SNe~IIP by \citet{anderson_characterizing_2014}. The depth of the dropoff is smaller than for the other SNe~II in comparison. \citet{valenti_diversity_2016} show that the subluminous explosions tend to have a greater dropoff depth, compared to the more typical SNe. SN~2021yja and SN~2004et have a similar behavior (slope) in their post-maximum plateau phase, yet they exhibit noticeably different dropoff slopes and $M_{\text{tail}}$ magnitudes. \citet{anderson_characterizing_2014} also show that a brighter tail-phase magnitude correlates with a higher $^{56}$Ni mass (see their Fig. 33). The long plateau and the shallow post-plateau dropoff indicate that the explosion of SN~2021yja is more energetic and synthesized more $^{56}$Ni compared to the average of SNe~IIP such as SN~1999em. This interpretation is also corroborated by the high photospheric velocity at early phases as inferred from spectroscopic modelling (see Sec. \ref{sec:energetics} and \ref{sec:photo_temp_vel}).

We estimate the nickel mass from the bolometric luminosity of the exponential tail using the method presented by \citet{hamuy_observed_2003}. Using their Equations 1 and 2, our values for the total extinction $A_V = 3.2\,E(B-V) \approx 0.22$\,mag, explosion epoch $t_0 = 59,465$, %redshift $z = 0.0053$,
distance $D \approx 21.8$\,Mpc, and applying the bolometric correction  BC $\approx 0.26$\,mag from \citet{hamuy_distance_2001}, we estimate the $^{56}$Ni mass of SN~2021yja to be $\sim 0.12\,M_{\odot}$ when measured at $t_{\rm tail} = 59,618.15$ with an apparent KAIT  $V_{\rm tail} \approx 16.94$\,mag. Doing the calculation for a few other points on the tail does not appreciably change the results. {For SN~1999em, \citet{hamuy_observed_2003} measured a $^{56}$Ni mass of $0.04^{+0.027}_{-0.019}\,M_{\odot}$.} \citet{misra_type_2007} determine an average $^{56}$Ni mass of $0.06 \pm 0.03\,M_{\odot}$ for SN~2004et using the same method at two different points on the tail. Our estimate is consistent with SN~2021yja having a higher bolometric luminosity on the exponential tail compared to {SN~1999em and} SN~2004et. {Compared with the mean $^{56}$Ni mass of $0.044\,M_{\odot}$ calculated for Type II SNe ($N=115$) by \citet{anderson_meta-analysis_2019}, our $^{56}$Ni measurement is greater by more than one standard deviation of their value.}

\subsection{Search for the RSG Progenitor}
\begin{figure*}
    \centering
    \includegraphics[width=0.8\textwidth]{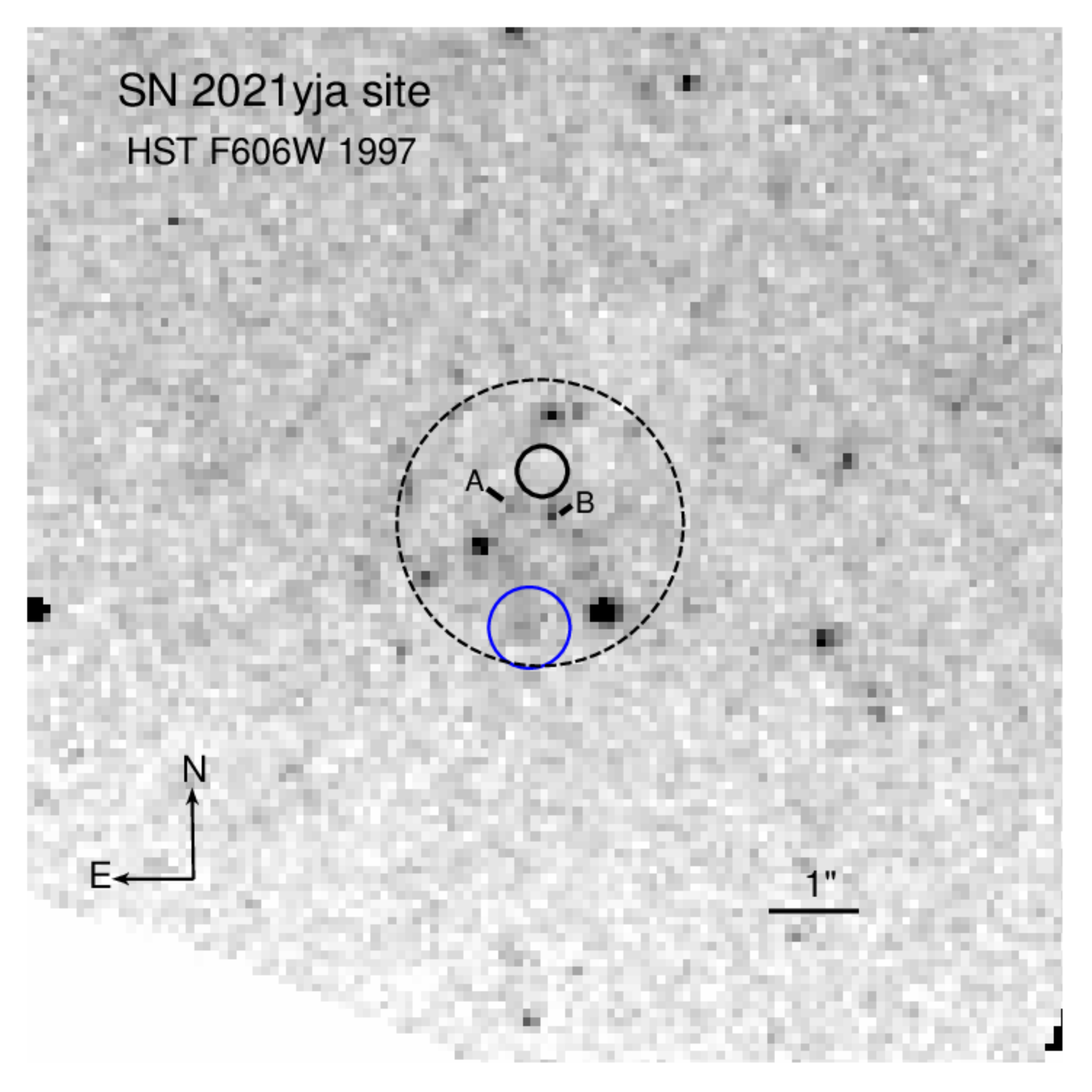}
    \caption{A portion of the pre-explosion {\sl HST\/} WFPC2 F606W image mosaic, obtained on 1997 March 26, containing the SN~2021yja site, is shown. 
    %The location of the SN was isolated via a relative astrometric match between the {\sl HST\/} mosaic and good-quality KAIT $R$-band images of the SN. 
    This SN location is indicated with a solid circle, with a radius of the $3\sigma$ formal astrometric uncertainty. The dashed circle encompasses the larger overall SN environment, which contains a notable clustering of several detected objects. Two stars that could be considered as progenitor candidates, if the progenitor had not experienced any circumstellar extinction, are indicated as ``A'' and ``B'' (see text). 
    {We also show with the solid blue circle the approximate position of the SN site as located by \citet{hosseinzadeh_weak_2022}.}
    %If the SN~2021yja progenitor were analogous to the SN~2017eaw progenitor \cite{dyk_type_2019}, it would not have been detectable in the F606W image.
    North is up, and east is to the left.}
    \label{fig:progenitor}
\end{figure*}

\begin{figure*}
    \centering
    \includegraphics[width=0.8\textwidth]{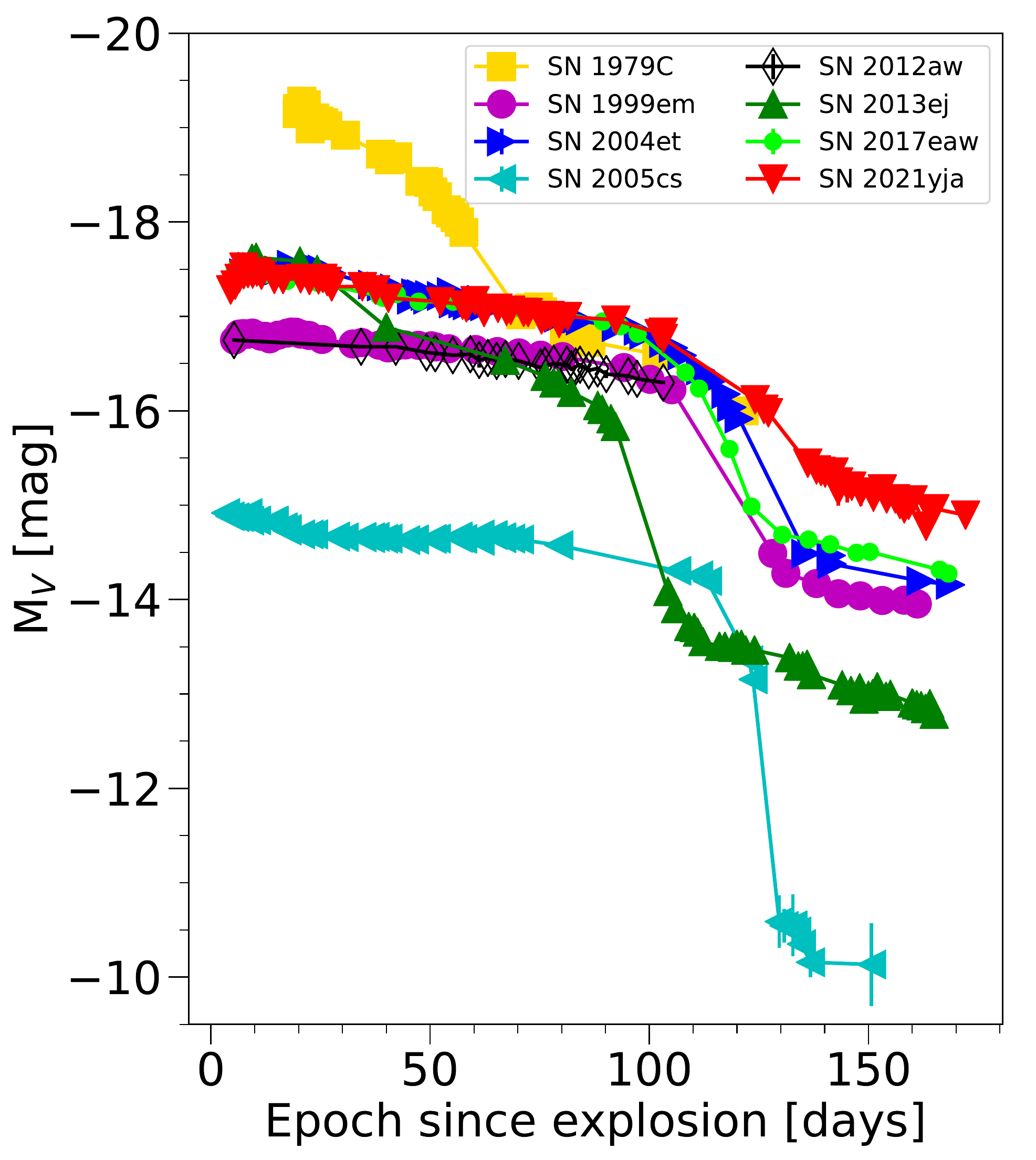}
    \caption{KAIT $V$-band light curve of SN~2021yja compared with that of other well-studied SNe~II: SN~1979C \citep{vaucouleurs_bright_1981}, SN~1999em \citep{leonard_distance_2002}, SN~2004et \citep{maguire_optical_2010}, SN~2005cs \citep{pastorello_sn_2009}, SN~2012aw \citep{dejaeger_berkeley_2019}, SN~2013ej \citep{huang_sn_2015},
    and SN~2017eaw \citep{dyk_type_2019}.} 
    \label{fig:lccomp}
\end{figure*}

%\textcolor{blue}
We investigated whether the SN progenitor was detected in pre-explosion {\sl HST\/} images available for the host galaxy in the archive. We found that there were two datasets, both obtained with the Wide Field and Planetary Camera 2 (WFPC2), one set on 1997 March 26 in the F606W filter (600\,s total exposure time) and the other on 2001 July 7 in the F450W and F814W filters (460\,s in each of the bands). Unfortunately, the latter did not contain the SN site. We attempted to isolate the location of the SN in the former, using a combination of two KAIT $R$-band images of the SN, both with comparatively good seeing. We were able to match 7 stars in common between the KAIT images and the WFPC2 F606W image mosaic, leading to a $1\sigma$ formal uncertainty in the relative astrometry of $\sim 1$ WFPC2 pixel. The SN location is shown in Figure \ref{fig:progenitor}. As can be seen, no source is detected at the SN location, to within the astrometric uncertainty.

%\textcolor{blue}
The SN site is contained within chip 2 of the WFPC2 F606W dataset. We employed Dolphot \citep{Dolphin2016} to perform photometry on that chip. To determine whether the SN progenitor had possibly been detected in the data, we made the following argument. If the progenitor of SN~2021yja were analogous to the progenitor of SN~2017eaw \citep{dyk_type_2019}, then its absolute magnitude should be $M_{\rm F606W} \approx -3.8$\,mag. It was shown for the latter progenitor star that substantial circumstellar dust must have been present prior to explosion. If this luminosity is also applied to the SN~2021yja progenitor, then, given our assumptions for the distance and total reddening, we would expect the apparent brightness of the star in the 1997 image to have $m_{\rm F606W} \approx 28.1$\,mag. However, we found that the detection threshold was $\sim 26.7$\,mag (formally at $3\sigma$), so the progenitor would not have been detectable. If the progenitor had not experienced circumstellar extinction, then we might expect the star to have $M_{\rm F606W} \approx -5.9$\,mag, for an initial mass of $12\,M_{\odot}$, or $-6.3$\,mag for $15\,M_{\odot}$ \citep{Stanway2018}. This would mean that the star might be detectable at $\sim 26.0$ and $\sim 25.6$\,mag, respectively. Two detected stars meeting these respective criteria are labeled in Figure \ref{fig:progenitor} as ``A'' and ``B,'' and could be considered progenitor candidates. Both stars can be seen outside our formal astrometric uncertainty, but are in the general vicinity of the SN location. Clearly, an image of the SN with higher spatial resolution, obtained with {\sl HST\/} or with adaptive optics {(AO)} from the ground, is required to pinpoint the SN's location in the pre-explosion data. (We know from \citealt{VanDyk2014} that the SN location determined from low-resolution, ground-based SN imaging can be significantly displaced from that determined via high-resolution imaging.) We attempted to use a coadded mosaic of the 3 very short (4-sec) MIRVIS exposures from our {\it HST}/STIS observations to do this; however, unfortunately, no other objects besides the SN were visible in that mosaic. {We note that \citet{hosseinzadeh_weak_2022} obtained an AO image and more precisely located the SN site (we show this location in Figure~\ref{fig:progenitor}), although they were unable to identify a progenitor candidate.}

\subsection{The UV Spectrum}
It has been suggested that SNe~IIP show remarkable similarities in their UV spectra (2000--3500\,\AA) at $\sim 10$\,days after the explosion \citep{gal-yam_galex_2008,bufano_ultraviolet_2009,dhungana_extensive_2016}. Here we investigate the early UV spectroscopic properties of SN~2021yja and test this claim through a thorough comparison between our observations and those of other SNe~II which have early-time UV spectra.

In Figure~\ref{fig:hst_uvo} we present the {\it HST}/STIS UV-optical spectra of SN~2021yja obtained on days +9, +14, and +21. A comparison between the earliest {\it HST}/STIS spectrum and the {\it Swift}/UVOT spectra at days +5 and +7 is given in Figure~\ref{fig:hst_swift_uv}. We identify prominent, blueshifted hydrogen Balmer lines and the characteristic \ion{Mg}{2} $\lambda$2798 absorption line in the {\it HST}/STIS spectra of SN~2021yja. These features are also generally present in other SNe~II such as SNe~1999em, 2005cs, 2005ay, and 2012aw \citep{baron_preliminary_2000,brown_early_2007,gal-yam_galex_2008,bufano_ultraviolet_2009,bayless_long-lived_2013}. The UV spectrum blueward of 3000\,\AA\ is dominated by superposed \ion{Fe}{2} and \ion{Fe}{3} lines, causing a significant amount of line blanketing \citep{dessart_quantitative_2005}. The emission peaks at these wavelengths correspond to regions of reduced line blanketing \citep{brown_early_2007}.  By day 21, we observe a relatively smooth and featureless continuum in the 2000--3000\,\AA\ region as line-blanketing effects strengthen in the cooling ejecta.

\begin{figure*}
    \centering
    \includegraphics[width=1.0\textwidth]{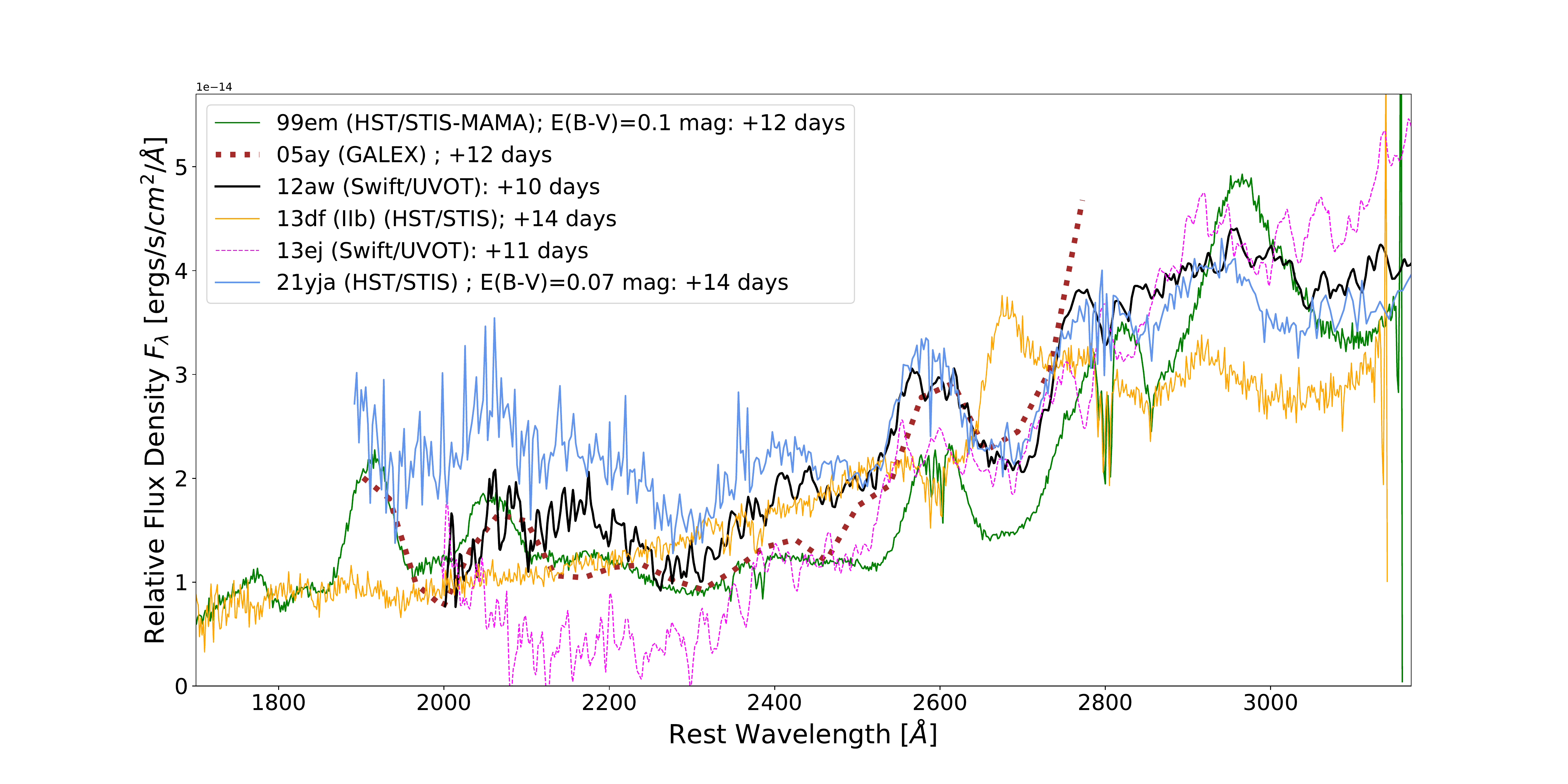}
    \caption{{\it HST}/STIS UV spectrum of SN~2021yja at +14\,days compared with the well-studied Type IIP SNe\,1999em \citep{baron_preliminary_2000}{, 2005ay \citep{gal-yam_galex_2008},} and 2012aw (\citet{bayless_long-lived_2013}; Rowe et al. 2022, in prep.), the Type IIP/L SN\,2013ej \citep{dhungana_extensive_2016} , and the Type IIb SN\,2013df \citep{ben-ami_ultraviolet_2015} observed at similar phases. All wavelength scales were corrected to the rest frame according to their host's redshift. All spectra were also dereddened assuming an $R_V = 3.1$ extinction law.}
    \label{fig:21yja_master_comp}
\end{figure*}

Figure~\ref{fig:21yja_master_comp} compares the UV spectrum of SN~2021yja at +14\,days with that of several other SNe~II observed at similar phases, including the Type IIP SNe\,1999em, {2005ay,} and 2012aw {\citep{baron_preliminary_2000,gal-yam_galex_2008,bayless_long-lived_2013}}, the transitional object SN~2013ej between Types IIP and IIL \citep{dhungana_extensive_2016},
and the Type IIb SN~2013df \citep{ben-ami_ultraviolet_2015}. {We include the last of these to demonstrate the effect of extreme CSM interaction on the UV flux.}
%In general, we observe homogeneity in the SN~IIP subclass 
In contrast to the previous suggestions of homogeneity, we see some similarities but also clear differences among the Type IIP UV spectra.
For example, SN~1999em shows a strong peak near 3000\,\AA\ that is not seen in SN~2021yja.
There are substantial differences among SNe~2012aw and 2013ej as well, %\textcolor{blue}
primarily at wavelengths redder than 2800\,\AA. {\citet{nagao_evidence_2021} found that SN 2013ej shows evidence for weak to moderate CSM interaction, which is consistent with the shallower P~Cygni trough and smoother UV flux.}
The SN~IIb~2013df appears to be quite different, with a prominent peak near 2700\,\AA\ where the other objects show a dip, and a more featureless continuum below 2600\,\AA.

We attribute the variations in the UV flux shape to a combination of factors, possibly including the epoch uncertainty and differences in metallicity, reddening, and pre-explosion behavior. 
%In the two {\it Swift}/UVOT spectra (SNe~2012aw and 2013ej), we observe much weaker features in the 2800--3200\,\AA\ range compared with the {\it HST} spectra of SNe~1999em, 2013df, and 2021yja.

%***\alex\Mazzali writes (and we can address now if there's time, later if no time now), "The MgII absorption changes shape quite dramatically between d9 and 14 (it develops a boxy trough). This may be worth a comment. Does the synthetic spectrum reproduce this behaviour, and if so what is the reason? Other lines? Ionization? Does it affect the width/velocity measurements mentioned later in the paper? [outside scope of paper]
The \ion{Mg}{2} $\lambda$2798 line is among the most well-studied features in the UV spectra of SNe~II. Its shape has been used to infer explosion properties of SNe~II (see Sec. \ref{sec:photo_temp_vel} and \ref{sec:energetics}). At +9 and +14\,days, the \ion{Mg}{2} $\lambda$2798 feature in SN~2021yja displays a prominent P~Cygni profile. By contrast, only a weak absorption component at $\sim 2700$\,\AA\ can be seen across the \ion{Mg}{2} $\lambda$2798 line observed at +5 and +7\,days (see Fig.~\ref{fig:hst_swift_uv}). %\textcolor{blue}{
The \ion{Mg}{2} absorption changes shape quite dramatically between +9 and +14\,days as it develops a boxy trough (Figure \ref{fig:mgii}), compared to H$\beta$ (Figure \ref{fig:hbeta}) and H$\alpha$ (Figure \ref{fig:halpha_hst}). By day +21, the P~Cygni profile overlaps with a series of metal lines and therefore appears less prominent. The broad P~Cygni feature is indicative of \ion{Mg}{2} formation close to the photosphere (see Sec. \ref{sec:originuv}). This is in stark contrast to SN~2013df, which shows an asymmetric and strongly blueshifted \ion{Mg}{2} line, suggesting that its formation is far above the photosphere and in the CSM \citep{ben-ami_ultraviolet_2015}. 

We also identify the \ion{Fe}{2} line at 2900\,\AA, commonly observed in the other SNe~II. Of note is the line blanketing by \ion{Ti}{2} and \ion{Ni}{2} around 3000\,\AA\ and 2500\,\AA, respectively. These lines of singly ionized species were also observed in SN~2005cs and discussed by \citet{bufano_ultraviolet_2009}. 
%\N{(so is that FeII emission at 2900/3000 A in 1999em?  that looks really striking.)} \\

The UV spectra of SN~2021yja are also distinct from those of peculiar Type II SNe such as SN~1987A, which displays a sharp cutoff in the UV flux below 3000\,\AA\ \citep{kirshner_ultraviolet_1987,pun_ultraviolet_1995}. There are also notable differences between the SNe~IIP and the interacting SN~IIb~2013df, primarily in the latter's blueshifted \ion{Mg}{2} P~Cygni feature and relatively smooth spectrum below 2600\,\AA\ \citep{ben-ami_ultraviolet_2015}.

\subsection{The Optical Spectrum}
The optical spectral evolution of SN~2021yja is presented in Figure \ref{fig:kast_hst}. We also compare the +6\,day Kast spectrum to that of other well-studied SNe in Figure \ref{fig:optical}, showing that the early optical spectrum fits within the framework. A total of 12 spectra were obtained by Kast and {\it HST}, spanning from +4 to +121\,days.  

Within the first $\sim 10$\,days after the SN explosion, the 
optical spectra of SN~2021yja are characterized by a series of 
broad Balmer lines superimposed on a hot, featureless continuum. 
We associate the weak feature around 5800\,\AA\ at +9\,days with 
the \ion{He}{1} $\lambda5876$ line. After $\sim +30$\,days, the 
\ion{O}{1} $\lambda$7774 and the \ion{Ca}{2} NIR triplet became 
progressively dominant at red wavelengths, corresponding to the 
plateau phase of the SN. The temporal evolution of the P~Cygni 
profile across the H$\alpha$ feature of SN~2021yja appears to be 
similar to those observed from other SNe~IIP 
\citep{gutierrez_type_2017}.
We do not observe a strong \ion{Si}{2} $\lambda$6355 line that was reported for the transitional Type IIP-IIL SN~2013ej at $\sim 2$--3 weeks \citep{valenti_first_2014}. 
The blue part of the optical spectra after day +30 is characterized by a series of features of intermediate-mass and heavy elements such as Na, Ca, and Fe. The P~Cygni profiles are also becoming more pronounced. 
%***\alex\Mazzali asks, "Any indication that there may be a contribution from HeI?"

We observe a change in the location of the H$\alpha$ emission peak over time. The evolution of the H$\alpha$ emission line is plotted in Figure \ref{fig:halpha} for two {\it HST} and four Kast epochs. The {\it HST} spectrum at +14\,days shows a blueshift of nearly $4000$\,km\,s$^{-1}$. The blueshift of the  emission component decreases roughly linearly with time, reaching a velocity of $\sim 1000$\,km\,s$^{-1}$ by day +61.  This blueshifted emission peak (a few thousand km\,s$^{-1}$) is consistent with a steep density profile and is commonly observed in other SNe~IIP \citep{dessart_quantitative_2005,anderson_analysis_2014} (see Sec. \ref{sec:density}). 

A notch forms in the left wing of the H$\alpha$ absorption at +121 days. This feature has been interpreted as evidence for high-velocity H$\alpha$ absorption caused by CSM interaction \citep{chugai_optical_2007}. However, we do not observe late-time asymmetries in the H$\alpha$ emission peak that were present in the spectrum of SN~2013ej at around 129\,days, which were interpreted as evidence for CSM interaction; see \citet{dhungana_extensive_2016}. 

Although SNe~IIL are spectroscopically similar, they show characteristically flatter Balmer lines compared to SNe~IIP, likely owing to a lack of hydrogen in deeper and slower-expanding layers of ejecta or to interaction with the CSM \citep{hillier_photometric_2019}. 
%anyone know of a good couple of papers to cite here?

\subsection{Modeling with TARDIS}
\label{s:tardis}

Type II SNe have been previously modeled using sophisticated radiative-transfer codes, notably by \citet{baron_type_2004} using \texttt{PHOENIX}, as well as by \citet{dessart_quantitative_2006} and  \citet{dessart_using_2008} with \texttt{CMFGEN}. 
For spectral modeling, we use a modified version of the Monte Carlo radiative-transfer code \textsc{TARDIS} developed for the analysis of SNe~II \citep{kerzendorf_spectral_2014,vogl_spectral_2019}. The code treats the excitation and ionization of hydrogen in 
non-local thermodynamic equilibrium (NLTE). The level populations of other elements are generally calculated with a simplified NLTE treatment --- the nebular approximation of \citet{1993A&A...279..447M} --- to reduce computational costs. This potentially limits the accuracy for estimates of species ionization and thus the metallicity.
%***\alex\Mazzali says, "Actually, the nebular approx used in ML93 was tested against a full NLTE code in Pauldrach et al 96, with very good agreement." Sergiy: This should probably asked to Christian V., since he added this extra piece in.

In our analysis of the {\it HST} spectra, we assume a power-law density profile and a homogeneous composition. The models differ in the steepness of the density profile $n$, temperature, velocity, metallicity, and time since explosion. In addition to the SN parameters, we vary the value for the host extinction, $E(B-V)_{\text{host}}$.
We use a machine-learning emulator, trained on a large grid of \textsc{TARDIS} simulations, to fit the models to the data as in \citet{vogl_spectral_2020}. The emulator generates synthetic spectra much faster for new parameter combinations than \textsc{TARDIS}, allowing us to find the best-fitting parameters through $\chi^2$ minimization. Only the third epoch falls within the parameter space of the emulator from \citet{vogl_spectral_2020}. We use a new set of models that extends to higher temperatures and velocities, and to steeper density profiles for the first two epochs. Compared to the older simulations, we treat helium in NLTE to produce more accurate spectra at high temperatures.
%\textbf{$<$ C.V.}

\begin{figure*}
    \centering
    \includegraphics[width=0.6\textwidth]{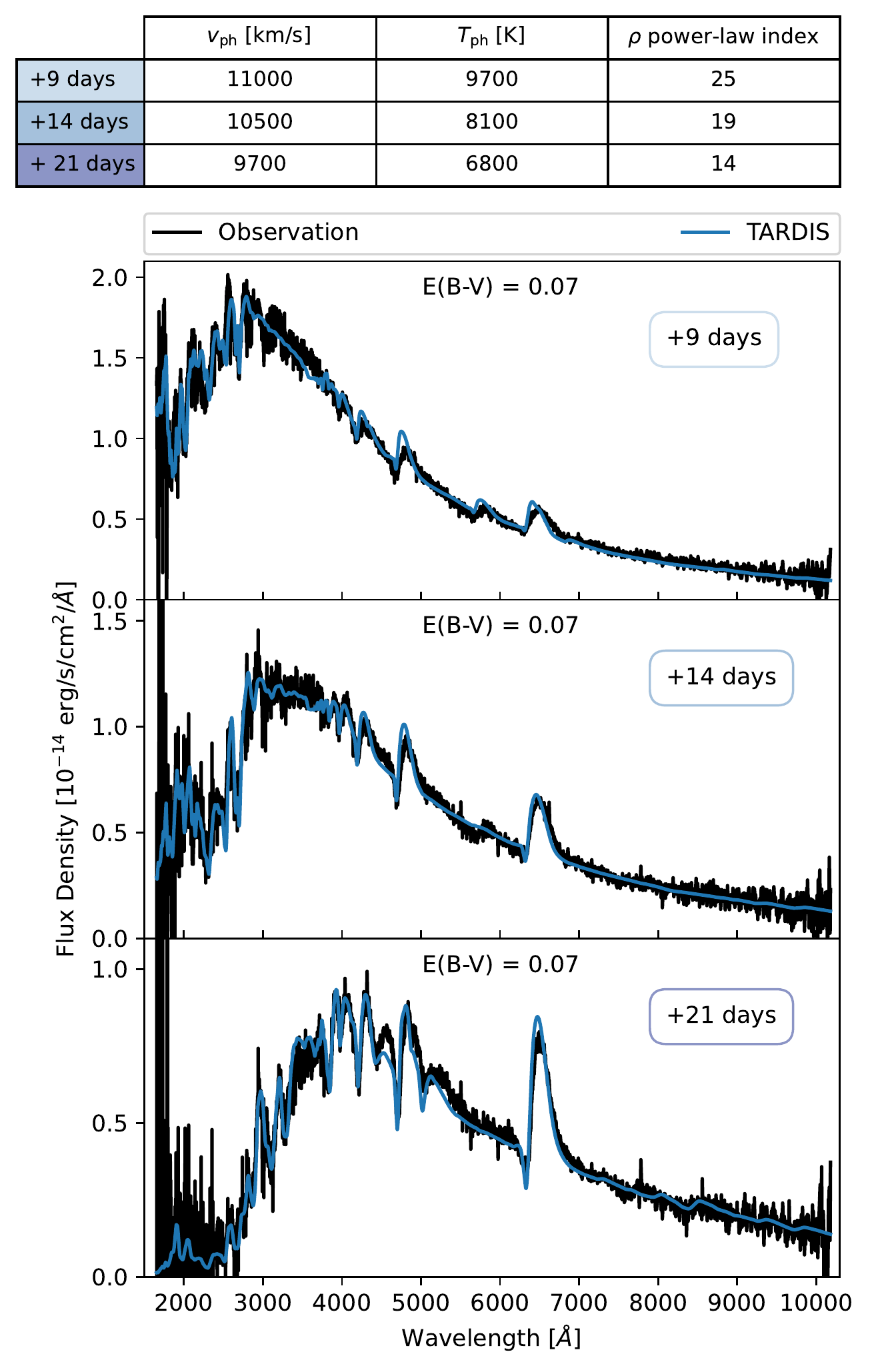}
    \caption{\texttt{TARDIS} fits to the {\it HST} STIS UV-optical spectra of SN~2021yja at +9 (top), +14 (middle), and +21 days (bottom). The best-fit parameters are presented above the top panel.}
    \label{fig:tardis_fits}
\end{figure*}

Our \texttt{TARDIS} fits for the three {\it HST} observations are shown in Figure \ref{fig:tardis_fits}. The plot includes a table with estimates of the key fit parameters: the photospheric velocity $v_\mathrm{ph}$, the photospheric temperature $T_\mathrm{ph}$ (i.e., the gas temperature at the photosphere), and the power-law index of the density profile $n$. The fits indicate some additional host-galaxy extinction $E(B-V)_\mathrm{total}\approx 0.07$\,mag and a subsolar metallicity.{ Given the uncertainties in the modeling (e.g., the approximate NLTE treatment of metal species)
and a certain degree of degeneracy between metallicity, temperature, and extinction, we provide only a qualitative estimate for the metallicity.}

\begin{figure*}
    \centering
    \includegraphics[width=1\textwidth]{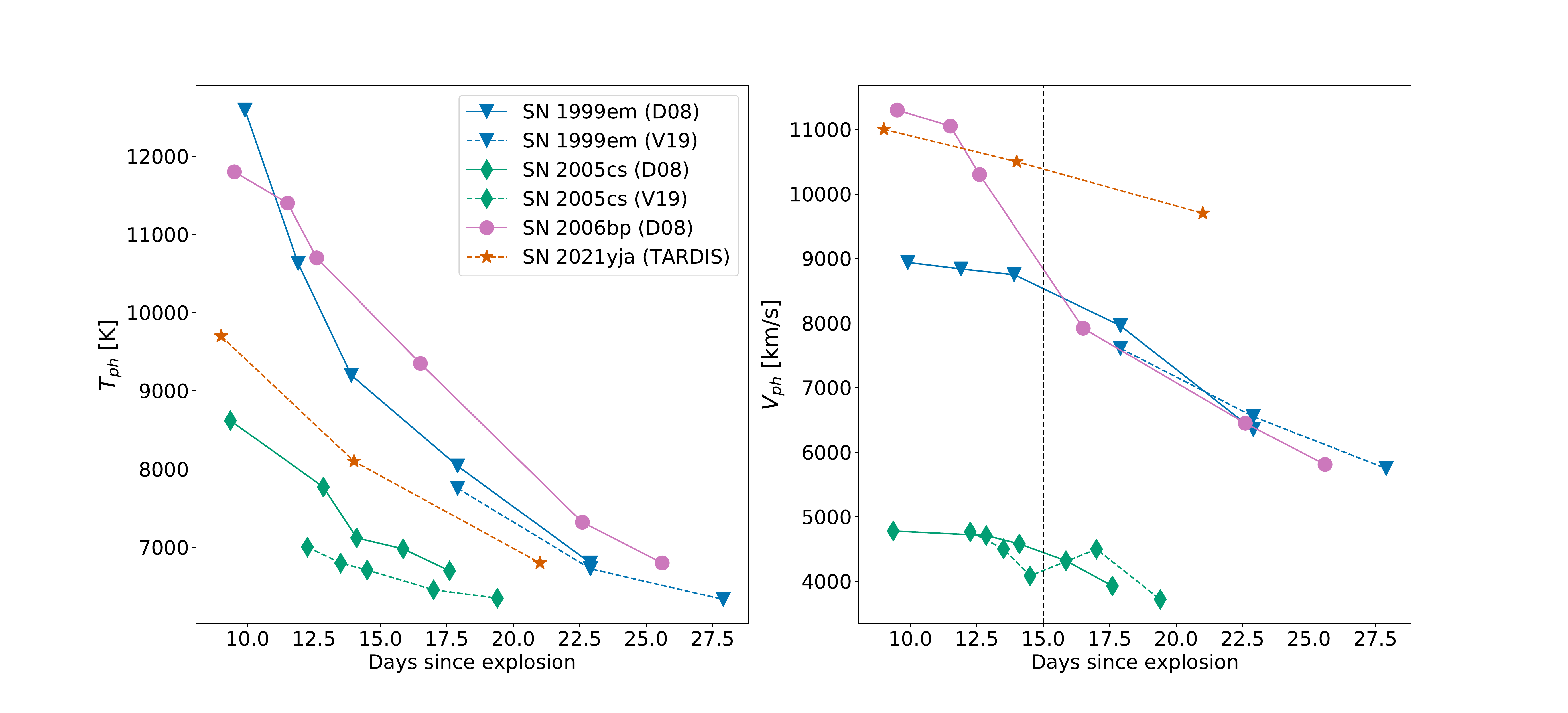}
    \caption{Photospheric temperature $T_{\rm ph}$ (left) and photospheric velocity $v_{\rm ph} $ (right) evolution of SN~2021yja obtained from the \texttt{TARDIS} fit to the three epochs of {\it HST}/STIS spectra compared to that of SNe~1999em, SN~2005cs, and SN~2006bp  (\citealp{dessart_using_2008}, D08; \citealp{vogl_spectral_2019}, V19). Vertical dashed line indicates the velocity at +15\,days, $v_{\text{d}15}$, for comparison purposes.}
    \label{fig:modeling_comp}
\end{figure*}

%\subsubsection{Chemical Abundances}
%-Include Ionization as function of radius if possible//
%-Include inferred density power law profile if possible
\subsubsection{Evolution of the Photospheric Temperature and Velocity}
\label{sec:photo_temp_vel}
%In this section, we discuss the inferred photospheric temperature $T_{\text{ph}}$ and velocity $v_{\text{ph}}$ for SN\,2021yja.
In Figure \ref{fig:modeling_comp}, we plot the time-series evolution of $T_{\text{ph}}$ and $v_{\text{ph}}$ for SNe~1999em, 2005cs, 2006bp, and 2021yja. We use the inferred values from \citet[][hereafter D08]{dessart_using_2008}, \citet[][B13]{bose_supernova_2013}, and \citet[][V19]{vogl_spectral_2019} in our comparison. The high photospheric velocities found for SN~2021yja are comparable to those of SN~2006bp (D08).

The photosphere cools as the ejecta expand adiabatically. The inferred $T_{\text{ph}}$ for SN~2021yja is consistent with the other SNe~IIP in our comparison. In our range of observations, we observe that $T_{\text{ph}}$ declines roughly linearly with time. SNe~IIP are shown to slow down their rate of cooling as they approach a temperature of $\sim 6000$\,K, indicating the onset of the % photospheric 
plateau phase. 
The photospheric temperature is roughly similar to that of SN~2012aw and SN~2013ej (not plotted here), while the subluminous SN~2005cs has $T_{\text{ph}}$ smaller by a factor of two.

\citet{faran_photometric_2014} find that the photospheric velocity is higher for SNe~IIL than for SNe~IIP. In the literature, the \ion{Fe}{2} absorption minimum (e.g., $\lambda$5169) is used as a reasonable estimate for the photospheric velocity for SNe~IIP \citep{hamuy_distance_2001,leonard_distance_2002,dessart_quantitative_2005}, whereas in this work and that of D08 and V19, the photospheric velocity is obtained using radiative-transfer modeling. We infer a remarkably high photospheric velocity for SN~2021yja, starting at 11,000\,km\,s$^{-1}$ on day +9, slowly decreasing to 10,500\,km\,s$^{-1}$ on day +14 and 9700\,km\,s$^{-1}$ on day +21. We show in our comparison to other SNe in \ref{fig:modeling_comp} that SN~2021yja has a relatively high $V_{\text{ph}}$, well above that of SN~1999em and SN~2005cs, even exceeding that of SN~2006bp at later epochs. This suggests a high explosion energy compared to a typical SN~IIP (see Sec. \ref{sec:energetics}). 

\subsubsection{Density Profile and P~Cygni Blueshift}
\label{sec:density}
The radial density profile of SNe~II can be modelled by a power law,
$\rho = \rho_{0}(r/r_{0})^{-n} $,  where $\rho_{0}$ represents the density at a corresponding characteristic radius $r_{0}$ and $n$ denotes the index of the power law. Early observations of SNe~II exhibit a steep density decrease with radius which gradually flattens with time. For SNe~IIP, the typical value for $n$ during the first two weeks after explosion can be upward of 20, dropping down to roughly 10 during the photospheric phase \citep{dessart_using_2008,vogl_spectral_2019}. 
% DONT FORGET TO CITE!!
For SN~2021yja, we find that the temporal evolution of $n$ is consistent with the literature. The inferred values for $n$ are shown in the right-most column of Figure~\ref{fig:tardis_fits}. 

\begin{figure}
    \centering
    \includegraphics[width=0.5\textwidth]{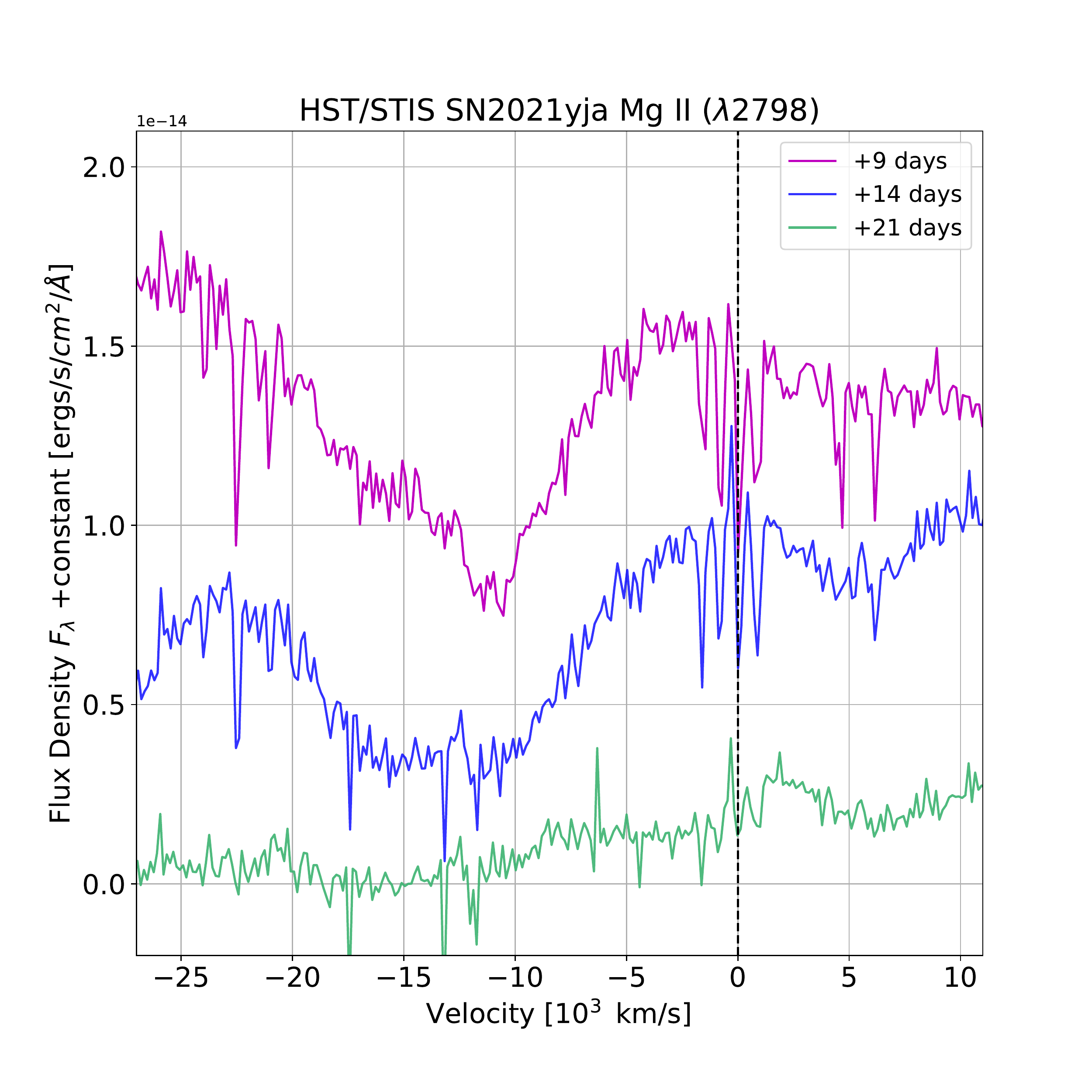}
    \caption{\ion{Mg}{2} line evolution for three {\it HST} epochs. Dereddened with $E(B-V) = 0.07$\,mag assuming $R_V=3.1$.}
    \label{fig:mgii}
\end{figure}

\begin{figure}
    \centering
    \includegraphics[width=0.5\textwidth]{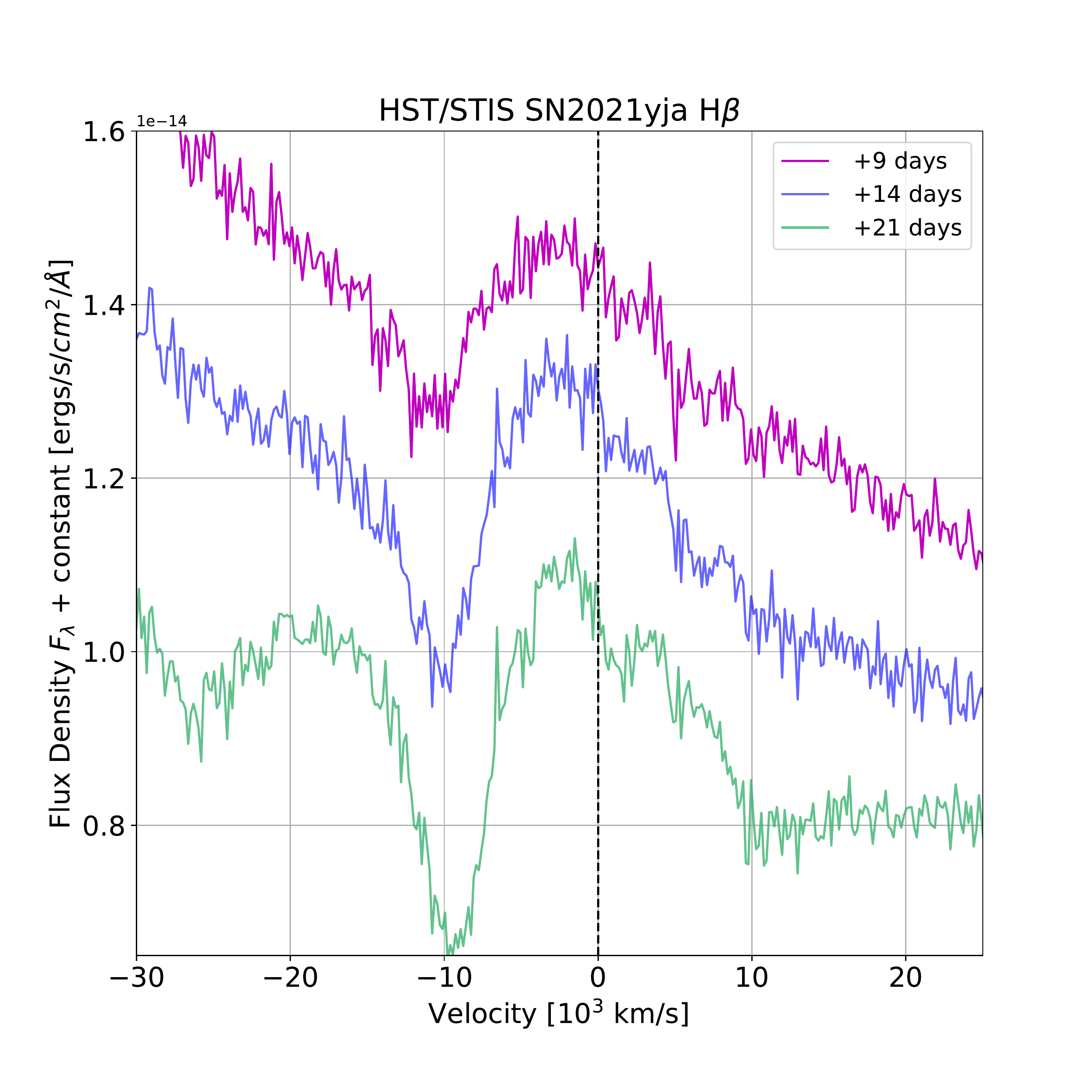}
    \caption{H$\beta$ line evolution for three {\it HST} epochs. Dereddened with $E(B-V) = 0.07$\,mag assuming $R_V=3.1$.}
    \label{fig:hbeta}
\end{figure}

\begin{figure}
    \centering
    \includegraphics[width=0.5\textwidth]{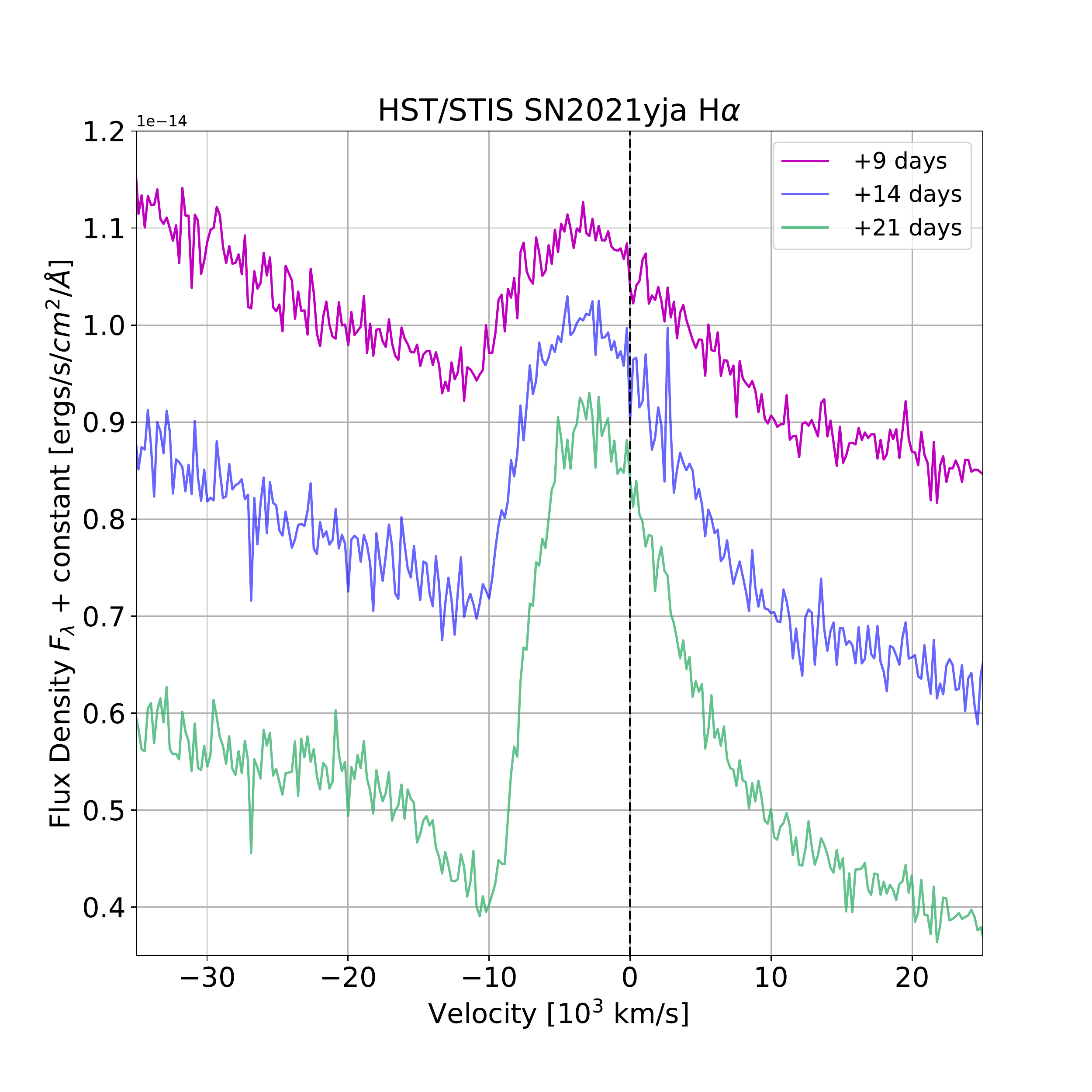}
    \caption{H$\alpha$ line evolution for three {\it HST} epochs. Dereddened with $E(B-V) = 0.07$\,mag assuming $R_V=3.1$.}
    \label{fig:halpha_hst}
\end{figure}

\begin{figure}
    \centering
    \includegraphics[width=0.5\textwidth]{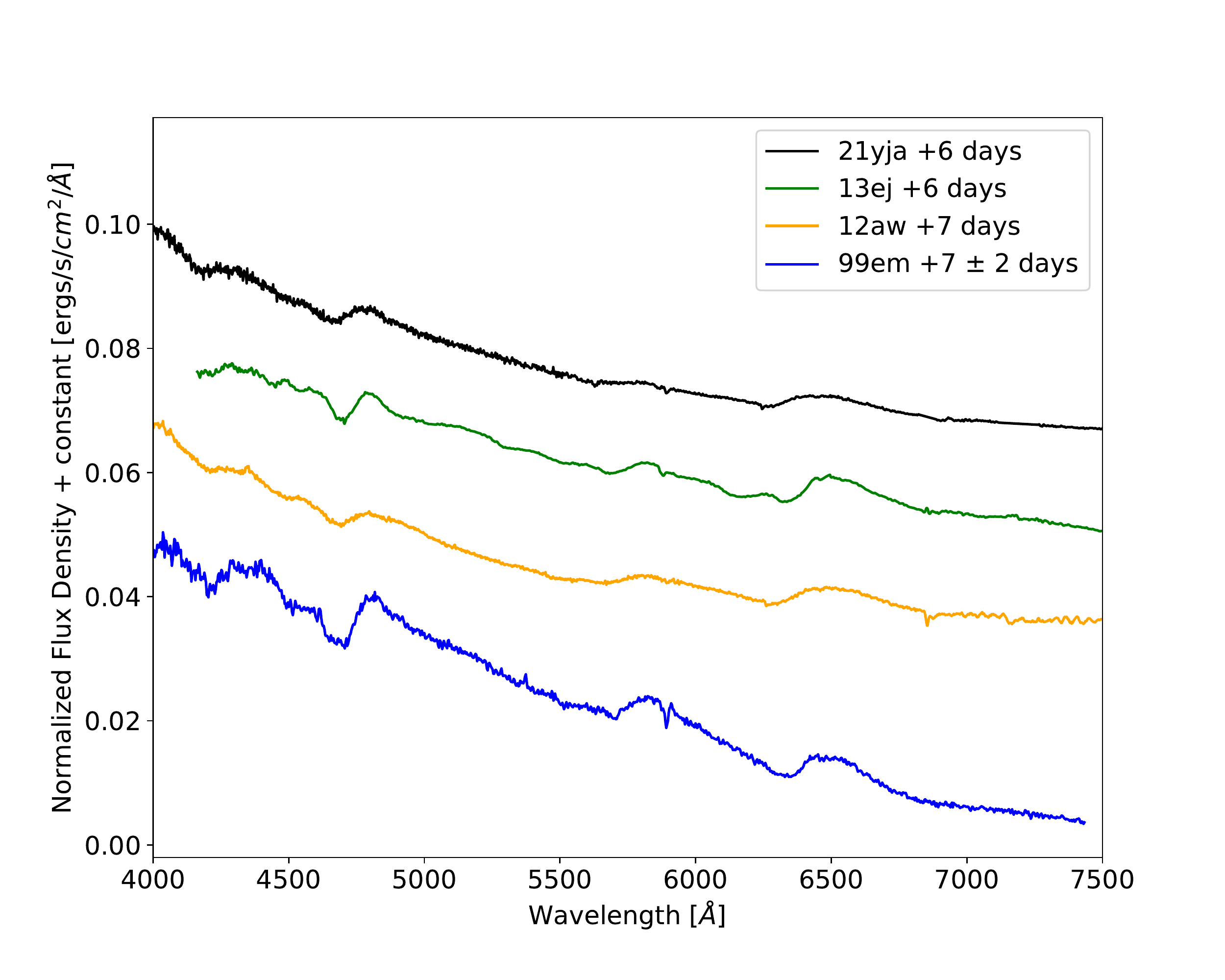}
    \caption{Comparison of SN~2021yja optical spectrum at +6\,days to that of other SNe~IIP.}
    \label{fig:optical}
\end{figure}

\begin{figure*}
    \centering
    \includegraphics[width=0.8\textwidth]{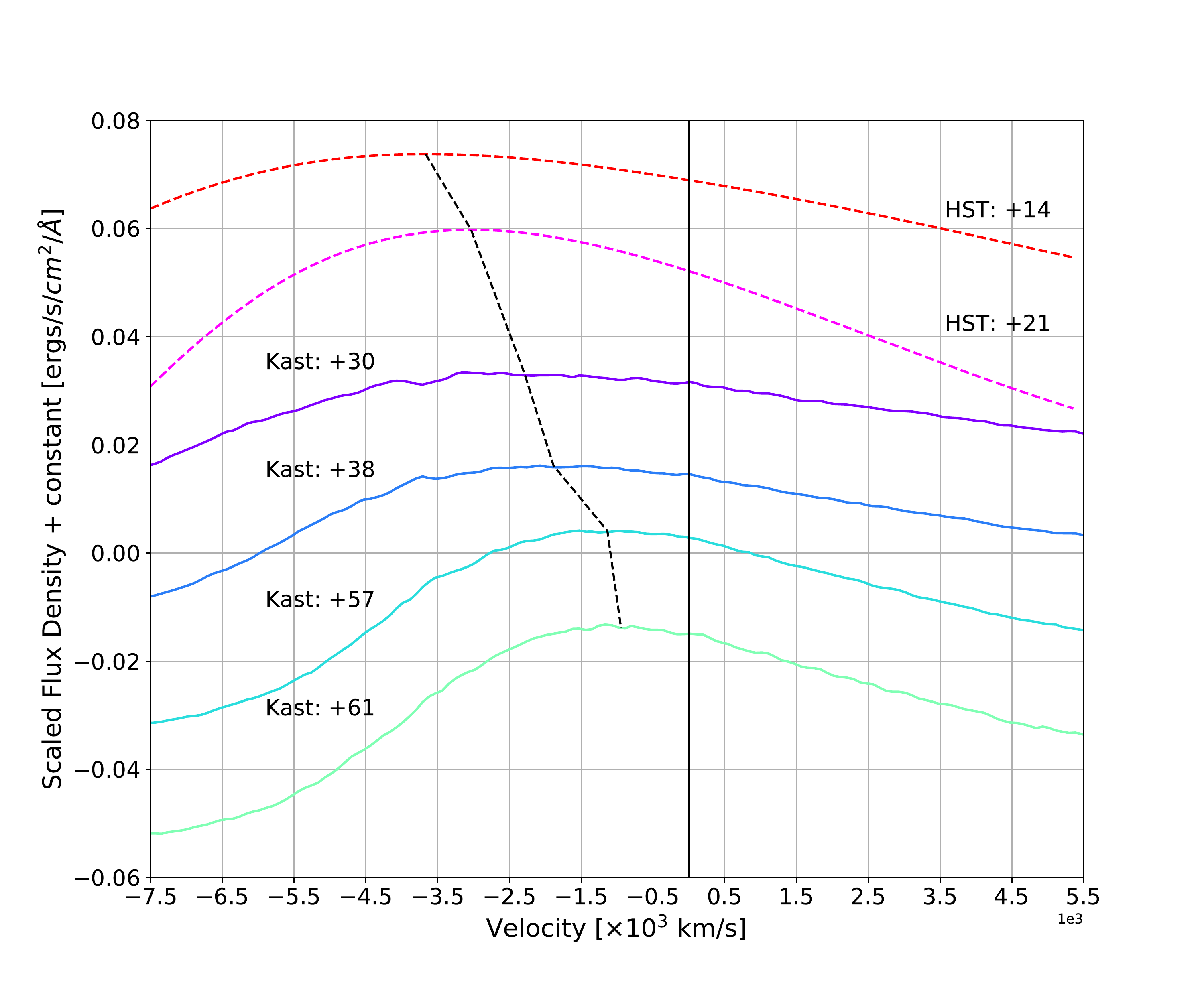}
    \caption{Evolution of the H$\alpha$ profile of SN~2021yja from days +14 to +61 presented in velocity space.  Color-coded, dashed curves present high-order polynomial fits to the {\it HST} spectra, which are used to estimate the location of the emission peaks. The black-dashed curve traces the shift of the H$\alpha$ emission peak. All spectra are corrected for the redshift of the host galaxy and for the extinctions from the host and the Milky Way. The spectra were scaled arbitrarily for clarity.}
    \label{fig:halpha}
\end{figure*}

A consequence of the varying density profile is the temporal blueshift of the H$\alpha$ emission peak \citep{anderson_analysis_2014}. We observe this effect in Figure \ref{fig:halpha}, with the H$\alpha$ emission peak weakening in its blueshift as the density profile power-law index falls from $n = 25$ on day +9 to $n = 14$ on day +21. Other effects of changing $n$ are discussed by \citet{dessart_quantitative_2005}, but are beyond the scope of this work.

\subsection{Energetics}

\label{sec:energetics}
The photospheric velocity serves as a useful tracer of SN kinematics. The principle of such a method --- in particular, a relationship between the kinetic energy of the SN ejecta and the photospheric velocity at 15\,days after shock breakout, $v_{d15}$ --- has been discussed by \citet{dessart_determining_2010} (see their Fig. 4). Explosions with an energy of 1\,foe ($10^{51}$\,ergs) have $v_{d15} < 9000$\,km\,s$^{-1}$, whereas explosions with an energy of 3\,foe have $v_{d15} > 12,000$\,km\,s$^{-1}$.
Although we do not have any data for the +15\,day phase, we can approximate $v_{d15}$ by using the photospheric velocity at +14\,days. SN~2021yja appears to be a high-energy/velocity clone of SN~1999em and closer to SN~2012aw (see \citealt{bose_supernova_2013}). 

Based on the \texttt{TARDIS} modelling from days +9 to +21, we estimate that SN~2021yja has $v_{\text{d}15} \approx 10,300$\,km\,s$^{-1}$, yielding an asymptotic ejecta kinetic energy of 1--3\,foe, whereas SN~1999em has $\sim 1$\,foe and SN~2005cs has $\sim 0.1$--0.3\,foe. SN 2021yja has an initial explosion energy higher than typical SNe~IIP, similar to SN~2006bp, considering that both SNe exhibit comparable photospheric velocity evolution at early times. 

\subsection{Origin of the UV and Optical Flux}
\label{sec:originuv}
 In this section, we discuss whether the mid-UV and optical flux originates from the photosphere or from CSM interaction. The \ion{Mg}{2} $\lambda$2800 line has been used to determine the origin of the UV flux for a small sample of SNe~II. For example, \citet{brown_early_2007} showed that the \ion{Mg}{2} $\lambda$2798 line in the SN~2005cs {\it Swift}/UVOT spectrum has a broad P~Cygni profile with velocity comparable to $v_{\rm ph}$, suggesting that the UV flux originates in the photosphere, not in CSM. On the other hand, \citet{pun_ultraviolet_1995} found that the highly interacting Type II-L  SN~1979C had a narrower \ion{Mg}{2} line in addition to a relatively smooth UV spectrum compared with SNe~IIP.
 A smooth UV spectrum that does not show any obvious broad absorption features within a few weeks after the explosion is expected for SNe that interact with thick circumstellar envelopes \citep{pun_ultraviolet_1995}. For example, the CSM-interacting Type IIb SN~2013df shown in Figure \ref{fig:21yja_master_comp} has a smooth mid-UV continuum, weak absorption lines, and a blueshifted \ion{Mg}{2} emission peak. Other examples include Type IIb SN~1993J, which also had a smooth UV spectrum lacking any broad absorption features, similarly to SNe~IILs~1979C and 1980K \citep{jeffery_hubble_1994}.

%"In contrast, the UV spectrum for SN
%1993J, a SN IIb, taken by FOS of HST about 18 days after
%the explosion does not show any obvious broad absorptions
%(Jeffery et al. 1994), and instead resembles SN 1979C and SN
%1980K." - - Pun et al. 1995
The detection of X-rays in the first weeks after the explosion has been used to argue for CSM interaction in SNe~II.  \citet{immler_x-ray_2007} and \citet{dessart_using_2008} showed that in the case of SN~2006bp, although X-rays were observed for up to 12\,days after explosion, suggesting that interaction between the ejecta and the CSM occurred, there was no sizeable contribution to the UV or optical flux. X-ray emission suggesting CSM interaction has also been observed for more-recent events such as SN~2012aw \citep{immler_swift_2012} and SN~2013ej \citep{margutti_swift_2013}. The modeling fits for SN~2021yja manage to reproduce both the UV and optical spectrum without needing to take into account an additional heating source, such as a ejecta-CSM interaction. In other words, the CSM does not provide a sizeable effect on the flux, unlike what is observed for the interacting SN~IIb~2013df (\citealt{ben-ami_ultraviolet_2015}; see their Fig.~5). It is therefore unlikely that the CSM is a significant contributor to the high continuum polarization of SN~2021yja observed during the photospheric phase (Vasylyev et al., in prep.) These conclusions support a photospheric origin for the UV flux. 

\citet{fransson_physical_1984} suggested that the UV flux below 1500\,\AA, which primarily contains emission lines of highly-ionized species (\ion{N}{5}, \ion{N}{3}, \ion{Si}{5}), may originate from the SN ejecta interacting with the pre-existing CSM. However, the wavelength coverage of the SN~2021yja observations does not capture this parameter space; thus, we limit our discussion to the origin of the near-UV to mid-UV flux, 1700--3200\,\AA.

The \ion{Mg}{2} velocity widths of SN~2021yja are comparable to those of H$\beta$, suggesting that the UV flux originates close to the photosphere, unlike in SNe highly interacting with CSM (see Figures \ref{fig:mgii} and \ref{fig:hbeta}). In summary, we do not see strong evidence for the existence of CSM that would significantly affect the UV and optical flux.

%\textbf{SPECTROPOLARIMETRY WILL NOW BE STANDALONE PAPER}

\section{Conclusions}\label{s:conc}
We present multi-epoch {\it HST}/STIS and {\it Swift}/UVOT UV spectra of the young, nearby Type II-P SN~2021yja. Kast optical spectra are also given for six epochs of the photospheric phase up until the plateau dropoff. We compare the UV/optical spectrum of SN~2021yja to that of previously studied SNe with high-S/N data at similar epochs.
SN~2021yja fits well within the framework of other SNe~IIP, primarily in the shape and location of the strong \ion{Mg}{2} P~Cygni profile. 

Using the \texttt{TARDIS} code, we infer useful parameters of the explosion, including photospheric velocity, photospheric temperature, the density profile power-law index, and the metallicity. We qualitatively show that the luminous tail phase and high photospheric velocity suggest that SN~2021yja underwent a more energetic explosion compared to SN~1999em-like SNe~II, also producing more $^{56}$Ni. 
We do not find evidence from our modeling for a significant contribution to the UV flux by CSM interaction. This and the high-energy nature of the explosion suggest an aspherical explosion, given that SN~2021yja was found to have high continuum polarization ($p \approx 0.8$\%) during the photospheric phase (Vasylyev et al., in prep.).

%\begin{appendix}

%\end{appendix}

\begin{acknowledgments} 

This research was funded by {\it HST} grants AR-14259 and GO-16178 from the Space Telescope Science Institute (STScI), which is operated by the Association of Universities for Research in Astronomy (AURA), Inc., under NASA contract NAS5-26555. Additional generous financial support was provided to A.V.F.'s supernova group at U.C. Berkeley by Steven Nelson, Landon Noll, Sunil Nagaraj, Sandy Otellini, Gary and Cynthia Bengier, Clark and Sharon Winslow, Sanford Robertson, the Christopher R. Redlich Fund, and the Miller Institute for Basic Research in Science (in which A.V.F. was a Miller Senior Fellow).
The research of Y.Y. is supported through the Bengier-Winslow-Robertson Postdoctoral Fellowship. 

A major upgrade of the Kast spectrograph on the Shane 3\,m telescope at Lick Observatory, led by Brad Holden, was made possible through gifts from the Heising-Simons Foundation, William and Marina Kast, and the University of California Observatories.
KAIT and its ongoing operation were made possible by donations from Sun Microsystems, Inc., the Hewlett-Packard Company, AutoScope  Corporation, Lick Observatory, the National Science Foundation, the University of California, the Sylvia \& Jim Katzman Foundation, and the TABASGO Foundation. Research at Lick Observatory is partially supported by a generous gift from Google. We thank the staffs at STScI (especially Weston Eck and Svea Hernandez) and Lick Observatory for assistance with the observations. U.C. Berkeley undergraduates Ivan Altunin, Raphael Baer-Way,
Michael May, Vidhi Chander, and Evelyn Liu obtained some of the Lick/Nickel data.

{M.W. acknowledges support from the NASA Future Investigators in NASA Earth and Space Science and Technology grant (80NSSC21K1849) and support from the Thomas J. Moore Fellowship at New York University.}
{M.M.'s research is supported in part by {\it Swift} GI program 1619152 (NASA grant 80NSSC21K0280), TESS GI program G03267 (NASA grant 80NSSC21K0240), and by a grant from the New York University Research Challenge Fund Program.}

This research made use of \textsc{tardis}, a community-developed software package for spectral
synthesis in supernovae \citep{kerzendorf_spectral_2014, kerzendorf_wolfgang_2022_6299948}. The
development of \textsc{tardis} received support from GitHub, the Google Summer of Code
initiative, and ESA's Summer of Code in Space program. \textsc{tardis} is a fiscally
sponsored project of NumFOCUS. \textsc{tardis} makes extensive use of Astropy and Pyne.
\software{{Astropy \citep{astropy:2013, astropy:2018}, TARDIS \citep{kerzendorf_spectral_2014,vogl_spectral_2019}, uvotpy \citep{kuin_uvotpy_2014}, DAOPHOT \citep{stetson_daophot_1987}, IDL Astronomy user's library \citep{landsman_idl_1993}, SOUSA pipeline \citep{brown_sousa_2014}, Pyne \citep{scopatz_pyne_2012-1}}}
\bigskip

\end{acknowledgments}
\bigskip
\newpage
\clearpage
\bibliographystyle{aasjournal}
\bibliography{2021yja}

\listofchanges
\end{document}